\documentclass[a4paper,fleqn,usenatbib,useAMS]{mnras}

\usepackage{newtxtext,newtxmath}

\usepackage{graphicx}	
\usepackage{amsmath}	
\usepackage{amssymb}	
\usepackage{multicol}        
\usepackage{bm}		
\usepackage{pdflscape}	
\usepackage[T1]{fontenc}
\usepackage{ae,aecompl}
\usepackage{newtxtext,newtxmath}
\usepackage{float}
\usepackage{epsfig}
\usepackage[caption =false]{subfig}
\usepackage{float}
\usepackage{enumitem}
\usepackage{scrextend}

\def\lsim{\mathrel{\rlap{\lower4pt\hbox{\hskip1pt$\sim$}}
    \raise1pt\hbox{$<$}}}                
\def\gsim{\mathrel{\rlap{\lower4pt\hbox{\hskip1pt$\sim$}}
    \raise1pt\hbox{$>$}}}                

\title[Gravitationally-lensed reionization-era analogs]{RELICS: Spectroscopy of gravitationally-lensed $z\simeq 2$ reionization-era analogs and implications for CIII] detections at $z>6$}

\author[Mainali et al.]{Ramesh Mainali$^{1,2}$\thanks{E-mail:rmainali@email.arizona.edu},
Daniel P. Stark$^{1}$,
Mengtao Tang$^{1}$,
Jacopo Chevallard$^{3}$,  \newauthor
 St\'{e}phane Charlot$^{4}$, 
Keren Sharon$^{5}$, 
Dan Coe$^{6}$, 
Brett Salmon$^{6}$, 
Larry D. Bradley$^{6}$, \newauthor
Traci L. Johnson$^{5}$, 
Brenda Frye$^{1}$, 
Roberto J. Avila$^{6}$, 
Sara Ogaz$^{6}$, 
Adi Zitrin$^{7}$,  \newauthor
Maru\v{s}a Brada\v{c}$^{8}$, 
Brian C. Lemaux$^{8}$,
Guillaume Mahler$^{5}$, 
Rachel Paterno-Mahler$^{9}$,  \newauthor
Victoria Strait$^{8}$, 
Felipe Andrade-Santos$^{10}$
\\
$^{1}$Steward Observatory, University of Arizona, 933 N Cherry Ave, Tucson, AZ,USA\\
$^{2}$Observational Cosmology Lab, Goddard Space Flight Center, Code 665, Greenbelt, MD 20771, USA\\
$^{3}$ Scientific Support Office, Directorate of Science and Robotic Exploration, ESA/ESTEC, Keplerlaan 1, 2201 AZ Noordwijk, The Netherlands \\
$^{4}$ Sorbonne Universit\'{e}, UPMC-CNRS, UMR7095, Institut d'Astrophysique de Paris, F-75014 Paris, France \\
$^{5}$Department of Astronomy, University of Michigan, 500 Church Street, Ann Arbor, MI 48109, USA\\
$^{6}$Space Telescope Science Institute, 3700 San Martin Drive, Baltimore, MD 21218, USA\\
$^{7}$Physics Department, Ben-Gurion University of the Negev, P.O. Box 653, Be'er-Sheva 8410501, Israel\\
$^{8}$Department of Physics, University of California, Davis, CA 95616, USA\\
$^{9}$WM Keck Science Center, 925 North Mills Avenue, Claremont, CA 91711, USA\\
$^{10}$Harvard-Smithsonian Center for Astrophysics, 60 Garden Street, Cambridge, MA 02138, USA\\
}

\date{}

\pubyear{2019}

\begin{document}
\label{firstpage}
\pagerange{\pageref{firstpage}--\pageref{lastpage}}
\maketitle

\begin{abstract}

Recent observations have revealed the presence of strong CIII] emission (EW$_{\rm{CIII]}}>20$~\AA) 
in $z>6$ galaxies, the origin of which remains unclear.  In an effort to understand the nature of these line emitters, 
we have initiated a 
survey targeting CIII] emission in gravitationally-lensed reionization era analogs identified in {\it HST} imaging of 
clusters from the RELICS survey.  Here we report initial results on four galaxies selected to  have 
low stellar masses (2-8$\times$10$^7$ M$_\odot$) and J$_{125}$-band flux excesses indicative of intense [OIII]+H$\beta$ emission (EW$_{\rm{[OIII]+H\beta}}$=500-2000~\AA), similar to what has been observed at $z>6$.
We detect CIII] emission in three of the four sources, with the CIII] EW reaching values seen in the reionization era (EW$_{\rm{CIII]}}\simeq 17-22$~\AA) in the two sources with the strongest optical line emission (EW$_{\rm{[OIII]+H\beta}}\simeq 2000$~\AA). We have obtained a {\it Magellan}/FIRE near-infrared spectrum of the strongest CIII] emitter in our sample, revealing gas that is both metal poor and highly ionized. Using photoionization models, 
we are able to simultaneously reproduce the intense CIII] and optical line emission for extremely young (2-3 Myr) and 
metal poor (0.06-0.08 Z$_\odot$) stellar populations, as would be expected after a substantial upturn in the SFR of a low mass galaxy.   The sources in this survey are among the first for which CIII] has been used as the primary means of redshift confirmation. We suggest that it should  be possible to extend this approach to $z>6$ with current facilities, using CIII] to measure redshifts of  objects with IRAC excesses indicating
 EW$_{\rm{[OIII]+H\beta}}\simeq 2000$~\AA, providing a 
method of spectroscopic confirmation independent of Ly$\alpha$.
 \end{abstract}


\begin{keywords}
Galaxy evolution -- high-z -- lensed--ISM--reionization
\end{keywords}


\section{Introduction}

Over the past decade, our view of galaxies in the early universe has been revolutionized by
deep infrared imaging campaigns conducted with the {\it Hubble Space Telescope (HST)}.
These surveys have led to the discovery of a large photometric sample of galaxies at $z>6$ (e.g.
\citealt{McLure2013,Bradley2014,Finkelstein2015, Bouwens2015b, Livermore2017}; see \citealt{Stark2016} for a review),
providing our first window on the sources thought to be responsible for the reionization
of intergalactic hydrogen (e.g. \citealt{Robertson2013,Bouwens2015,Stanway2016,Finkelstein2019}).
Analysis of the broadband spectral energy distributions (SEDs) associated with these galaxies reveals a population of galaxies that are compact and low stellar mass, with large specific star formation rates (sSFR; e.g. \citealt{Labbe2013,Stark2013,Gonzalez2014,Salmon2015,Ono2013,Curtislake2016}). 

The first insights into the emission line properties have emerged from {\it Spitzer}/IRAC broadband 
photometry, revealing significant flux excesses in bandpasses covering rest-optical lines \citep{Ono2012,Finkelstein2013,Labbe2013,Smit2014a,Smit2015}.  The amplitude of the 
flux excess indicates the presence of extremely large equivalent width (EW) rest-frame optical line 
emission in some individual cases at $z>6$ (EW$_{\rm{[OIII]+H\beta}}\simeq 1000$-2000~\AA), as 
would be expected for galaxies dominated by very young ($\lsim 10$ Myr) stellar populations.  In other 
galaxies, older stellar populations ($\gsim 200$ Myr) appear present, and the [OIII] emission lines are less prominent 
(e.g., \citealt{Hashimoto2018,Strait2019}).  The typical reionization-era galaxy appears 
somewhat between these two extremes.  Analysis of composite SEDs constructed from galaxies 
at $z\simeq 7$  \citep{Labbe2013} reveals a stellar population that powers strong [OIII]+H$\beta$ emission 
(EW$_{\rm{[OIII]+H\beta}}\simeq 670$~\AA) but is also old enough (50-200 Myr) to have a 
small Balmer Break.  While the nebular emission implied by the average SED is less intense than 
in the individual examples noted above, it is still well above the threshold used to define 
extreme emission line galaxies (EELGs) at lower redshifts (EW$_{\rm{[OIII]\lambda5007}}>100$~\AA; \citealt{Amorin2014}),
 assuming standard [OIII]/H$\beta$ ratios for EELGs  \citep{Tang2019}. 
While these EELGs are rare at lower redshifts (e.g. \citealt{Atek2011, vanderwel2011}), they  
become more frequent at $z>6$ as  large sSFRs become typical \citep{Salmon2015}.  

In another few years, our view of these early star forming sources will be advanced by
the spectroscopic capabilities on the {\it James Webb Space Telescope (JWST)},
providing our first chance to study the nature of the massive stars and the metallicity of the ionized gas
in the reionization era.  A preview of the type of spectra that {\it JWST} is likely to observe at $z>6$
has recently emerged, revealing nebular emission line properties very different from what is common
at lower redshifts.   Strong CIII] emission has been identified in
galaxies at $z=6.03$ \citep{Stark2015a}, z=7.51 \citep{Hutchison2019}, 
and $z=7.73$ \citep{Stark2017}. 
The rest-frame CIII] EW is greater than 20~\AA\ in two of the three $z>6$ galaxies, more than an order of magnitude
greater than what is seen typically at $z\simeq 1-3$ (e.g. \citealt{Shapley2003,Du2017,Du2018,Steidel2016, LeFevre2019}). 
Strong nebular CIV emission has been detected in two low mass $z>6$ galaxies  \citep{Stark2015b, Mainali2017,Schmidt2017},
pointing to a hard ionizing spectrum that is usually associated with active galactic nuclei (AGNs) in galaxies at lower
redshifts \citep{Hainline2011,LeFevre2019}.  Most recently, the nebular NV emission line has been
reported in several  additional $z>7$ galaxies \citep{Hu2017,Tilvi2016,Laporte2017,Mainali2018}, requiring an
intense radiation field with significant flux of photons more energetic than 77 eV. Many 
of these sources have IRAC colors that  are suggestive of extreme optical line emission (EW$_{\rm{[OIII]+H\beta}}\simeq 1000$-2000~\AA), larger than the typical values inferred at $z\simeq 7$. 

The origin of the strong UV nebular line emission that we are seeing in $z>6$ galaxies remains unclear.  While the detection of nebular NV
likely points to AGN activity, the powering mechanism of the intense ($\rm{EW_{CIII]}}>20$~\AA)  CIII] emission
is still a matter of debate in the literature.  \citet{Stark2017} investigated
the spectral properties of the $z=7.73$ CIII] emitter using BEAGLE \citep{Chevallard2016},  a flexible tool for modeling
stellar and nebular emission in a consistent manner.   With the latest version of the \citet{Bruzual2003} population synthesis code (Charlot \& Bruzual 2019, in prep) as the input  radiation field, they found
that the SED and CIII] emission could be reproduced by models with the hard ionizing
spectrum of a metal poor ($\simeq 0.1-0.2$ Z$_\odot$) stellar population. Similar results were found in \citet{Stark2015a}
when fitting the spectrum of the $z=6.03$ CIII] emitter described above.  A  different picture
emerges from \citet{Nakajima2018}.  They explore the range of UV nebular line spectra that 
can be powered by stellar populations, considering both single star models using 
POPSTAR models \citep{Molla2009} and binary star 
 models using BPASS (v2, \citealt{Stanway2016}). Their results show that stellar photoionization 
is unlikely to power CIII] emission with $\rm{EW}>20$~\AA, only reaching 
such large values when C/O ratios are elevated above the solar abundance ratio, 
or if the stellar population is hotter than predicted by models as might be expected for an 
extremely top-heavy IMF.  As these cases are not likely to be the norm, they 
suggest that CIII] emitters with $\rm{EW}>20$~\AA\ are more likely explained 
by an ionizing spectrum from a mixture of young massive stars and AGN.

The tension between these two  interpretations highlights the challenges we are soon to face
once {\it JWST} begins collecting large spectroscopic samples at $z>6$.   The problem largely
reflects how little we know about the extreme UV (EUV) radiation field powered by low
metallicity massive stellar populations in high redshift galaxies.  Current population synthesis
models make different predictions for the shape of the EUV radiation field at low metallicity,
making it difficult to link the observed nebular line detections to a unique physical picture.  If
not addressed prior to the launch of {\it JWST}, there are bound to be large systematic
uncertainties in the physical properties derived from spectra at $z>6$.

Motivated by this shortcoming, attention has begun to focus on characterizing the rest-UV spectra
of star forming galaxies at $z\simeq 1-3$ (e.g. \citealt{Erb2010,Stark2014,Rigby2015, Du2017,LeFevre2019, Amorin2017}) and at $z< 0.01$ (e.g., \citealt{Berg2016, Berg2018, Senchyna2017, Senchyna2019}).   
These studies have demonstrated that prominent metal line emission does
appear in star forming galaxies, provided they are both low metallicity and have
a stellar population weighted toward very young ages.  The latter
trend is clearly seen in the relationship between CIII] EW and [OIII] EW. In spite of the 
strong CIII] emission, many of the metal poor line emitters are found to have 
sub-solar C/O ratios \citep{Erb2010,Berg2016,Berg2018}.
These observations can  be collectively explained by photoionization models 
(e.g. \citealt{Gutkin2016,Jaskot2016, Feltre2016, Byler2018}),
with intense UV metal line emission requiring the hard radiation field from young low metallicity stars and
the high electron temperature (T$_{\rm{e}}$) associated with metal poor gas.

Of particular interest is the nature of sources with extremely intense CIII] emission (EW$_{\rm{CIII]}}>20$~\AA),
matching the values seen in the first samples at $z>7$.  At intermediate redshifts ($z\simeq 1-3$), 
there are a small number of galaxies above this threshold  \citep{LeFevre2019, Amorin2017}, but they appear very 
different from the galaxies described above, with signatures of low luminosity AGN
or atypically large C/O ratios. These results are fully consistent with the picture put forward 
in \citet{Nakajima2018}, whereby the stellar radiation field is in most cases incapable of powering the intense 
line emission we are now observing at $z>7$.  If true, this would imply that low luminosity
AGN or large C/O ratios are  present among reionization-era spectroscopic samples.

However such conclusions remain premature as our current census of star forming galaxies is incomplete. 
While many metal poor galaxies have been targeted with rest-UV spectroscopy, most lack the extreme
optical line emission (EW$_{\rm{[OIII]+H\beta}}\simeq 2000$~\AA) which appears associated with the 
UV line emitters at $z>7$. The extremely young stellar populations (1-5 Myr) probed by these 
galaxies should produce a very intense EUV ionizing spectrum \citep{Tang2019}, yet little is known about the 
rest-UV spectra that they power.  To address this issue, we have initiated a campaign which
aims to characterize the range of CIII] EWs powered by metal poor galaxies with
large EW optical line emission. In order to efficiently identify EELGs, we apply well-established color
selection techniques \citep{vanderwel2011,Maseda2014} to pick out $z\simeq 1.6-1.8$ galaxies
with flux excesses in the J$_{\rm{125}}$-band from strong [OIII]+H$\beta$ emission.   To ensure our
sample preferentially consists of metal poor systems, we focus our search on gravitationally lensed
galaxies with blue UV colors and low masses in the Reionization Lensing Cluster Survey (RELICS) fields  \citep{Coe2019}. 
The combination of depth and volume probed by RELICS makes it ideal for building samples of low mass sources with
apparent magnitudes bright enough for ground-based spectroscopy.  In this paper, we present rest-UV spectra of 
four lensed EELGs (including two with EW$_{\rm{[OIII]+H\beta}}\simeq 2000$~\AA), allowing us to test whether the upper bound of the CIII] EW distribution in star forming galaxies extends to EW$_{\rm{CIII]}}>20$~\AA.  
We have obtained a rest-optical spectrum for the most extreme UV line emitter, allowing
us to investigate the origin of the line emission in more detail. 

This paper is organized into the following sections. 
We describe our photometric selection method and spectroscopic observations in \S2. 
We then discuss the  results of our spectroscopic observations in \S 3 and describe 
the ionized gas conditions and stellar population properties implied by the spectra in \S4. 
We then discuss implications for reionization-era
studies in \S5 and close with a brief summary in \S 6.
Throughout the paper, we adopt a $\Lambda$-dominated, flat universe
with $\Omega_{\Lambda}=0.7$, $\Omega_{M}=0.3$ and
$\rm{H_{0}}=70\,\rm{h_{70}}~{\rm km\,s}^{-1}\,{\rm Mpc}^{-1}$. We use solar oxygen abundance of 12+log(O/H)$_{\odot}$=8.69 \citep{Asplund2009}.
We quote magnitudes in the AB system and equivalent widths in the rest-frame.
  
  \begin{figure}
\centering
\hbox{\hspace{0 cm} \subfloat{\includegraphics[scale=0.45]{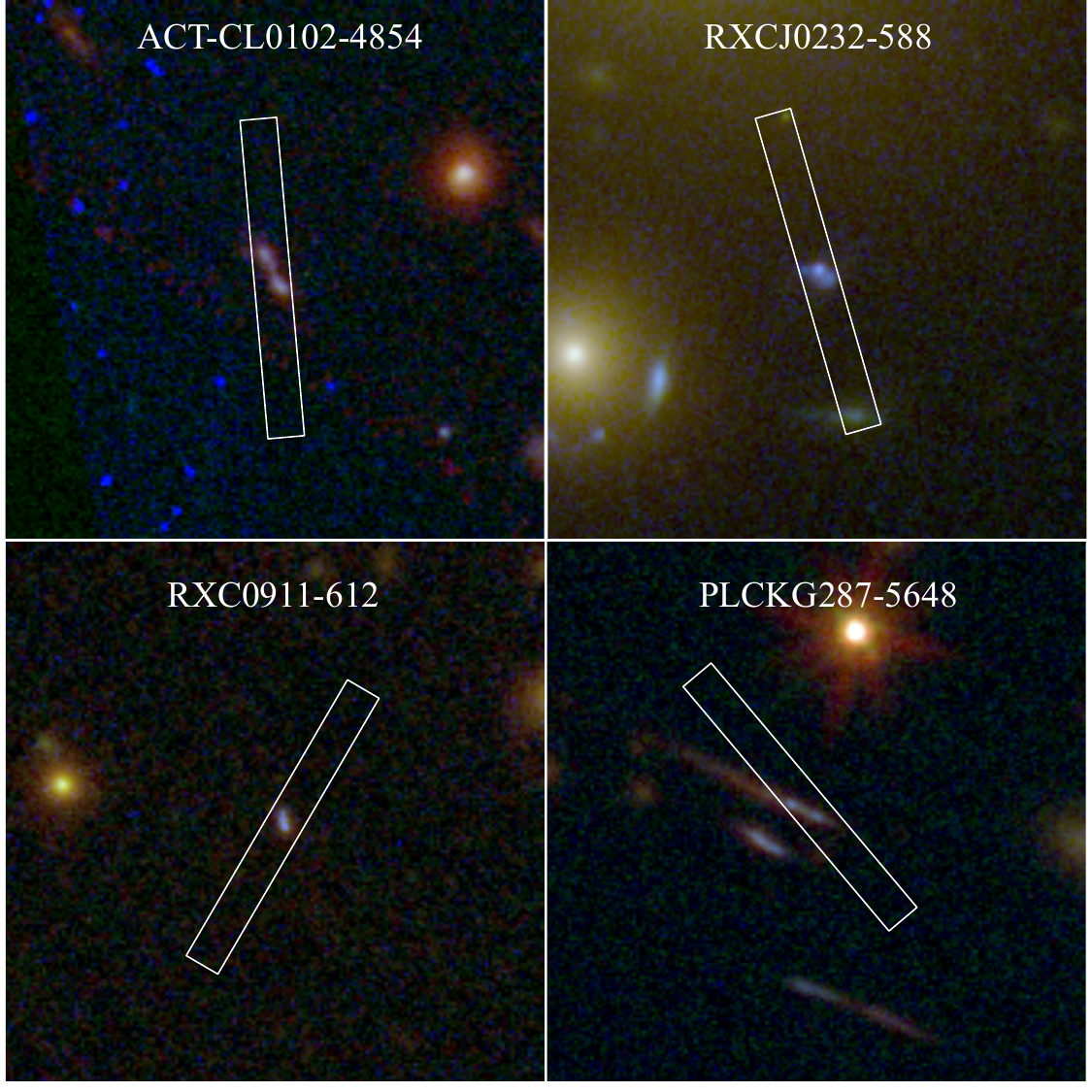}}}
\caption{ {\it HST} color images  of 
the lensed EELGs galaxies presented in this paper. The top of each image shows the cluster name followed by the galaxy name. The size of each image is 10\farcs$\times$10\farcs.}
\label{fig:sample_image2}
\end{figure}

\begin{figure*}
\centering
\includegraphics[scale=0.3,angle=0]{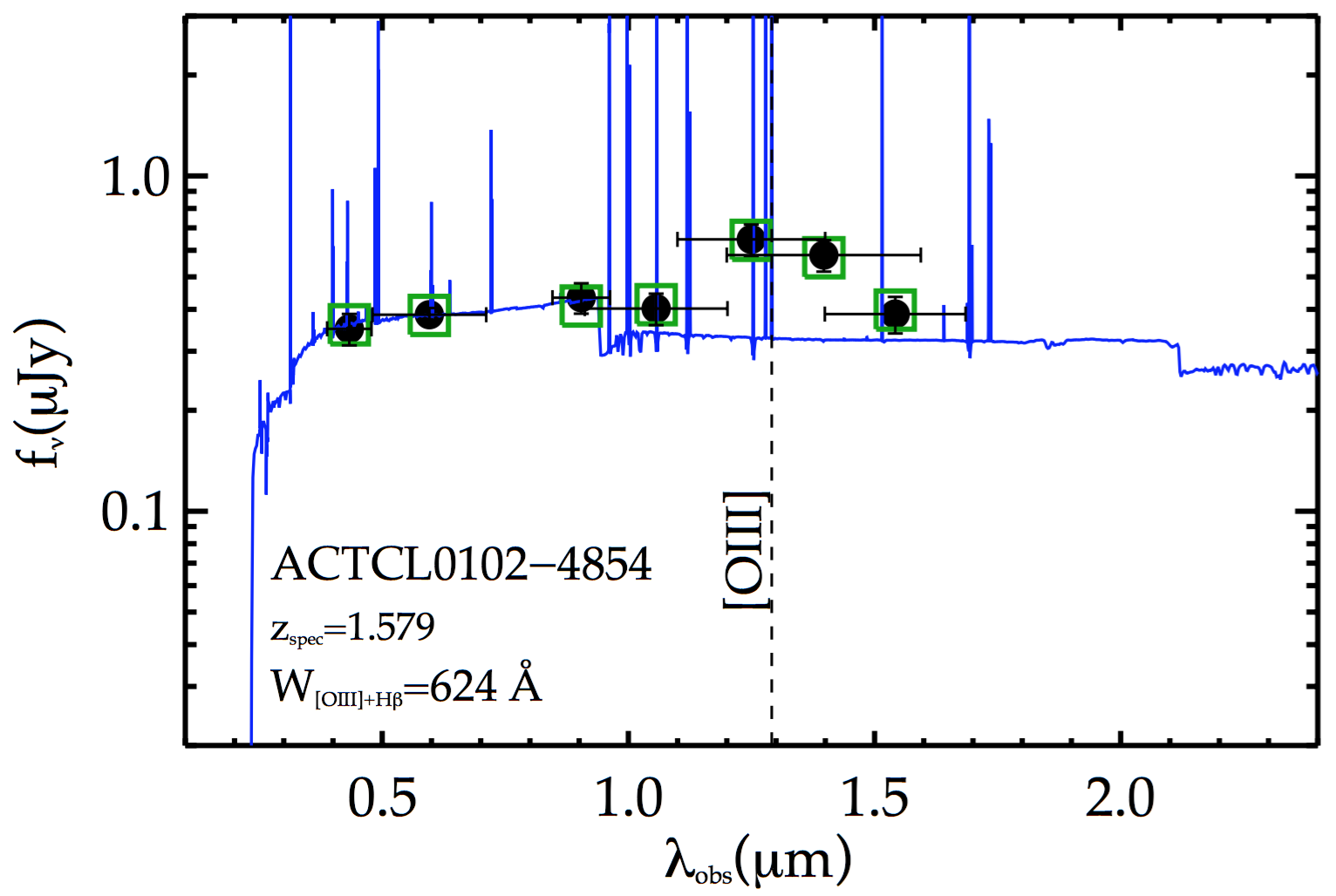}
\includegraphics[scale=0.3,angle=0]{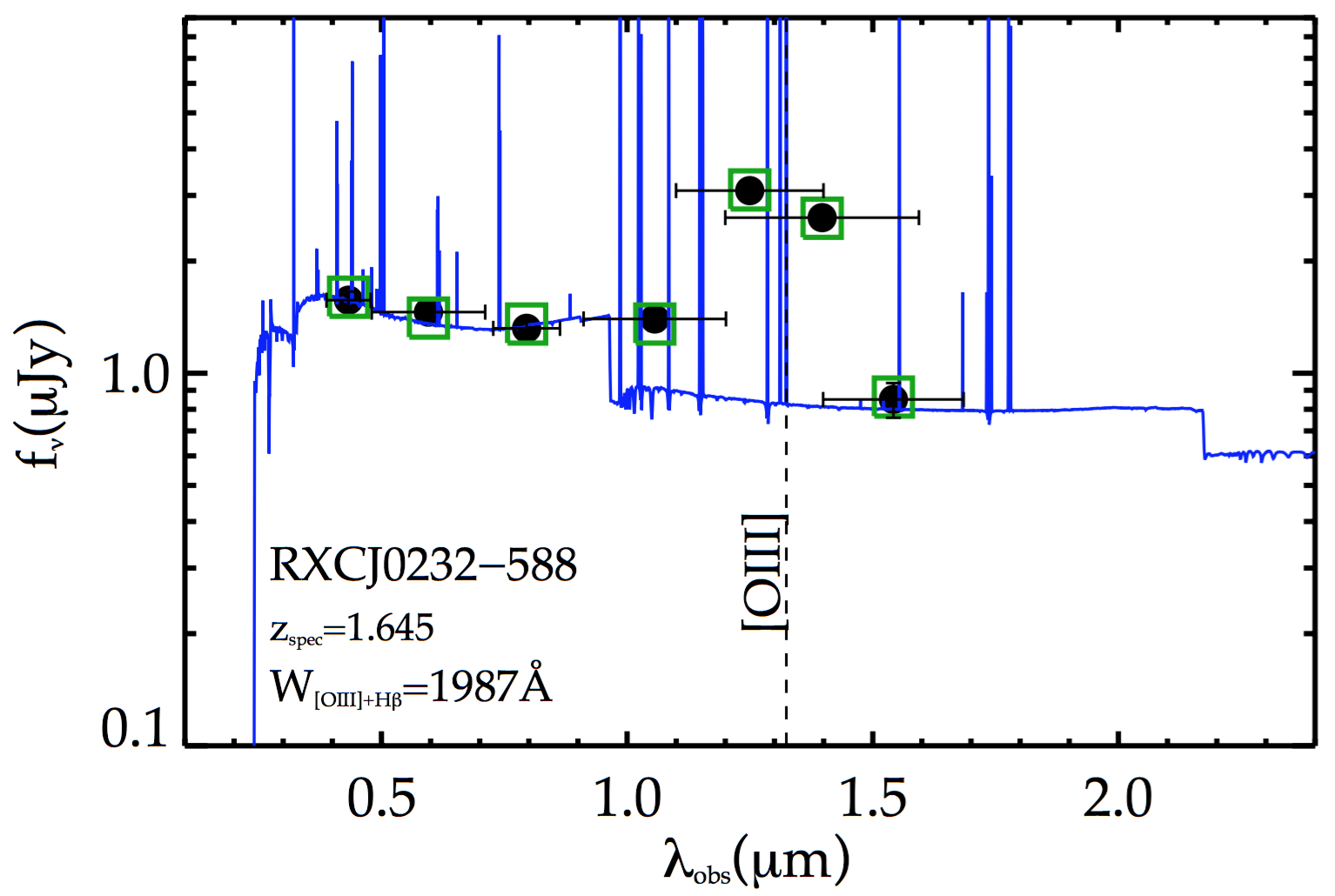}\\
\includegraphics[scale=0.3,angle=0]{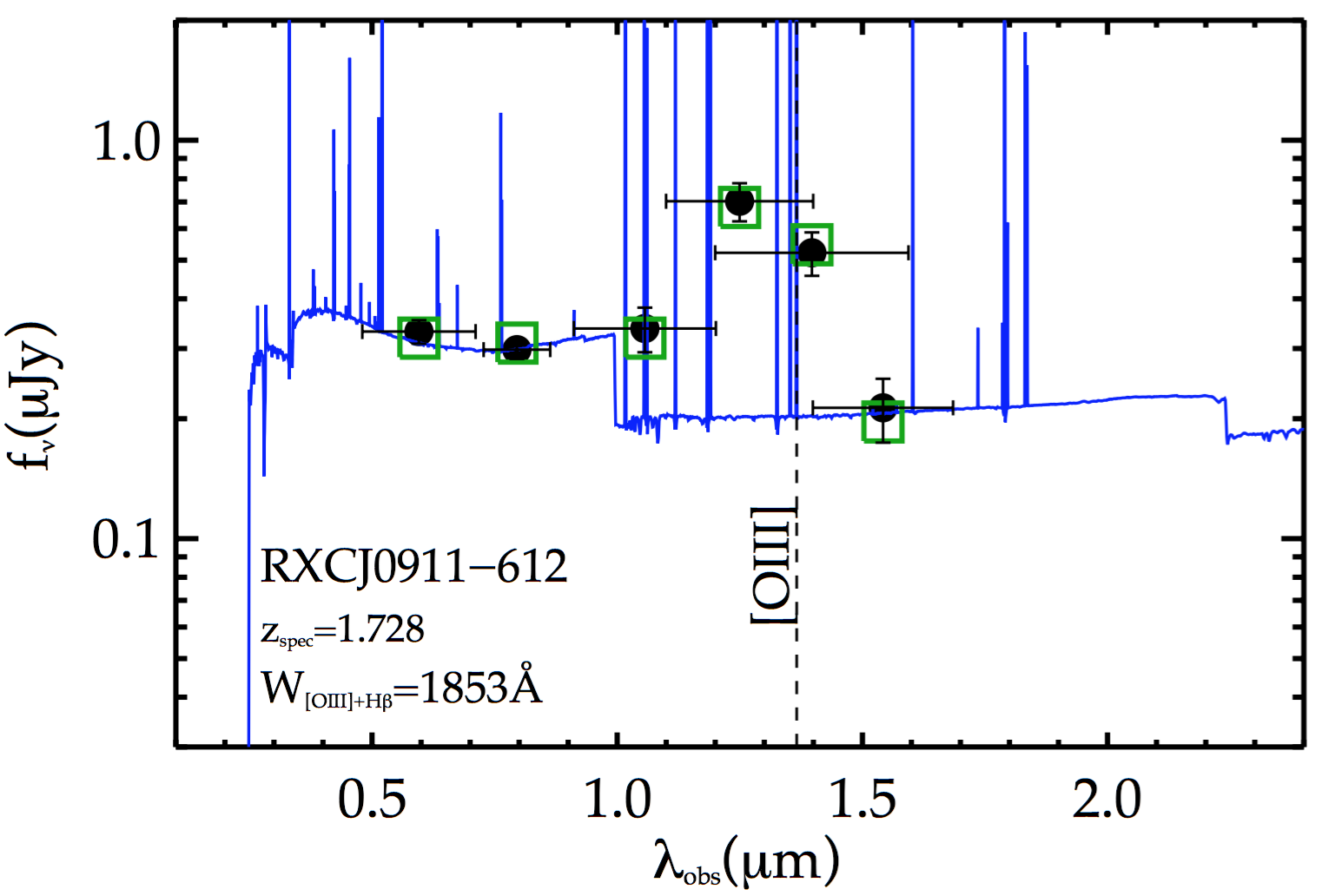}
\includegraphics[scale=0.3,angle=0]{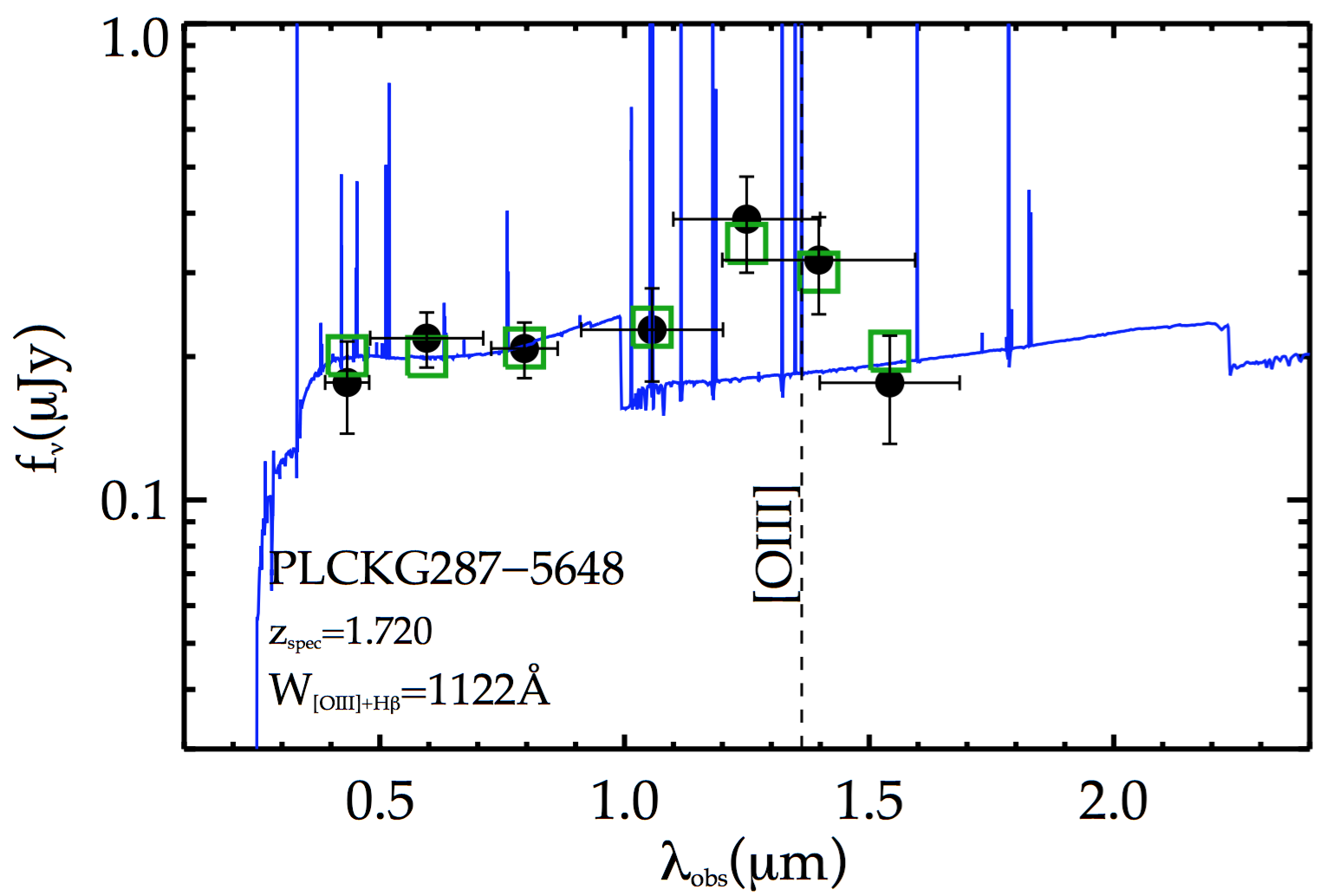}
\caption{Spectral energy distributions of four EELGs in our sample. The broad band flux excesses can be seen in the F125W and the F140W filters. 
The black circles and green squares represent observed and synthetic data points respectively. 
The blue curves show the best fit SED models inferred using BEAGLE (see \S 4.2). The bottom left of each panel shows the object ID, spectroscopic redshift and [OIII]+H$\beta$ equivalent width as implied by the J$\rm_{125}$-band
 flux excess.} 
\label{fig:sed}
\end{figure*}

\section{Spectroscopic Observations}\label{sec:observation}

\subsection{Selection of $z\simeq 2$ EELGs in RELICS fields}
Our goal is to characterize the range of rest-UV spectra  seen in galaxies with moderately low metallicities
and extreme EW optical line emission.   Recent studies have focused on $z\simeq 2$ EELGs in the CANDELS fields
(e.g., \citealt{Maseda2014,Tang2019}).   Here we seek to extend these efforts to lower mass and hence lower
metallicity EELGs.  We thus focus our search on gravitationally lensed galaxies in the 41 cluster fields imaged
by {\it HST} as part of the RELICS program (GO-14096;\citealt{Coe2019}).  The volume covered by this large dataset is ideal for
identifying highly-magnified objects that are bright enough for ground-based spectroscopy (m$_{\rm{AB}}<25.5)$. 

We first use {\it HST} imaging data to photometrically select our sample. 
The RELICS program provides {\it HST}/WFC3 infrared imaging data in four filters (F105W, F125W, F140W, and F160W filters) in all 41 clusters. 
For those clusters lacking existing archival ACS imaging, RELICS also provides
 {\it HST}/ACS optical imaging in F435W, F606W, and F814W filters (see \citealt{Coe2019} for details). In addition to RELICS program, the data used in this 
 paper comes from the online HST archive: GO-12477 (F. High), GO-12755 (J. Hughes), GO-9722 (H. Ebeling), GO-10493 (A. Gal-Yam), GO-10793 (A. Gal-Yam), GO-14165 (S. Seitz). 
We then make use of the publicly-available photometric catalog prepared by the RELICS team based on detection images in a weighted sum of ACS and WFC3 images\footnote{\label{footnote1}https://archive.stsci.edu/prepds/relics/}.

To efficiently select EELGs in the RELICS fields, we make use of well-established color selection criteria 
which identify the presence of strong [OIII]+H$\beta$ emission in galaxies at $z\simeq 1.57-1.79$
\citep{vanderwel2011,Maseda2014, Tang2019}.  At these redshifts, both lines are
situated in the J$_{\rm{125}}$-band (F125W).  Galaxies with large EW [OIII]+H$\beta$ emission are
easily identifiable by the large flux excess that appears in the J$_{\rm{125}}$-band relative to the 
adjacent I$_{\rm{814}}$ (F814W) and
H$_{\rm{160}}$ (F160W) bands.  We follow the color cuts used in \citet{vanderwel2011}, selecting sources with
$I_{814}-J_{125}>0.44+\sigma(I_{814}-J_{125})$ and $J_{125}-H_{160}<-0.44-\sigma(J_{125}-H_{160})$  where $\sigma$ refers to the uncertainty in the color.  
The amplitude of the  J$_{\rm{125}}$-band excess can be directly related to the rest-frame equivalent
width given by 
\begin{eqnarray} \rm{EW} = \Big(\frac{f_{total}-f_{cont}}{f_{cont}}\Big)
  \frac{\rm{W}}{1+z} \end{eqnarray} 
  where W is
the effective width of the J$\rm_{125}$ filter, f$\rm_{total}$ is the flux density
f$\rm_\nu$ in the J$\rm_{125}$ filter, and f$\rm_{cont}$ is the average flux density
in the I$\rm_{814}$ and H$\rm_{160}$ filters.  The color cuts described above thus select sources with
rest-frame [OIII]+H$\beta$ EW $>$ 500~\AA.

We apply the EELG selection criteria in all 41 RELICS clusters, utilizing the broadband fluxes from
the publicly-available photometric catalog prepared by the RELICS team.
Most cluster fields have 1-2 EELGs with typical apparent magnitude of $m\rm_{AB}\sim25.5$ in the I$\rm_{814}$ band.
We focus spectroscopic follow-up on sources that have bright continuum magnitudes ($m\rm_{AB}$<25.5) and
large J$_{\rm{125}}$-band excesses implying extremely intense optical line emission. 

In this paper, we present observations of four lensed EELGs from the parent sample with  600~\AA$<$ [OIII]+H$\beta$ EW $<$ 2000~\AA .
{\it HST} imaging postage stamps are shown in Fig.~1, revealing the very compact sizes of each of the four sources. 
The impact of [OIII]+H$\beta$ emission on the J$_{\rm{125}}$-band  is clearly visible in the
spectral energy distributions (SEDs) shown in Fig.~2.  We provide details of the four sources in
Table 1 and discuss each individually below.  The magnification provided by the cluster is calculated from lens models prepared
by the RELICS team and  available to the public.\footref{footnote1} Further
details on the lens modeling procedure have been presented in recent RELICS papers \citep{Cerny2018, PaternoMahler2018,
Cibirka2018, Acebron2018}.  In calculating the magnification and the absolute magnitude, we
assume the confirmed spectroscopic redshift presented in \S3. 

RXCJ0232-588 is a compact (r$\rm_{e}$=$0\farcs1$), blue ($\beta=-2.3$) galaxy with a
pronounced J$_{\rm{125}}$-band flux excess (Fig.~2) indicative of strong rest-optical nebular emission 
(EW$\rm_{[OIII]+H\beta}$=1990$\pm$200~\AA). 
Its apparent optical magnitude is the brightest in our sample (I$_{\rm{814}}$=23.6), making it an ideal target for spectroscopy. 
The RXCJ0232.2-4420 lens model indicates  a magnification factor
of $\rm \mu=7.96_{-1.72}^{+0.02}$.  After correcting for magnification,  we
calculate an absolute UV magnitude of M$_{\rm{UV}}=-18.72_{-0.02}^{+0.28}$.

RXCJ0911-612 is a bright (I$_{\rm{814}}$=25.2) compact  galaxy (r$\rm_{e}$=$0\farcs1$) with a prominent  $J_{125}$-band flux excess
 (Fig.~2), indicating extremely strong optical line emission ([OIII]+H$\beta$ EW = 1850$\pm$170~\AA).
Based on the lens model for RXCJ0911.1+1746, we estimate  that the cluster has magnified this
object by a factor of $\rm \mu=3.8_{-0.13}^{+0.11}$.   After correcting for the magnification, we
find that RXCJ0911-612 has an  absolute magnitude of M$\rm_{UV}$=-$17.92_{-0.03}^{+0.02}$.

PLCKG287-5648 was previously identified as a multiply-imaged system by \citet{Zitrin2017} (ID number 2 in
their paper).  Among the three images, only 2.3 has near-infrared imaging coverage necessary for the EELG color selection.
The source is relatively bright (I$_{\rm{814}}$=24.4), has a blue UV slope ($\beta=-1.9$),
 and shows a slightly extended arc-like structure in the {\it HST} images (Fig.~ 1).
The $J_{125}$-band flux excess is evident in the SED (Fig.~2), implying an [OIII]+H$\beta$ EW of 1120$\pm$150~\AA.
Using the cluster lens model, we find that the galaxy is magnified by a factor of $\rm \mu=13.6_{-3.2}^{+0.4}$. 
After correcting for this factor, we derive an absolute magnitude of M$\rm_{UV}$=-$17.14_{-0.02}^{+0.29}$, more than an
order of magnitude less than  M$^{\star}_{\rm{UV}}$ at $z\simeq 2$ (e.g., \citealt{Alavi2016}).

ACTCL0102-4854 is the second brightest source in our sample (I$_{\rm{814}}$=24.8) but has a less pronounced
$J_{125}$-band flux excess than the three sources discussed
above.  While still an EELG, its optical line EW is the weakest in our spectroscopic
sample (EW$\rm_{[OIII]+H\beta}$=620$\pm$220~\AA).  The source shows a slightly elongated structure (Fig.~ 1)
with a blue UV slope ($\beta=-1.9$). We compute a magnification factor of $\rm \mu=1.41_{-0.12}^{+0.12}$  using the RELICS lens
model for ACT-CL0102-49141.   Applying this magnification factor to the apparent magnitude, we
compute an absolute UV magnitude of M$\rm_{UV}$=-$19.84_{-0.14}^{+0.14}$, making ACTCL0102-4854 the
most luminous object in our sample. 

 \begin{table*}
\caption{Properties of gravitationally lensed EELGs presented in this paper.  From 
left to right, the columns denote the cluster name, object ID (taken from RELICS catalog), $\rm{z_{spec}}$ , RA and DEC of object,  
I$\rm_{814}$ band magnitude, UV slope, [OIII]+H$\beta$ equivalent widths (rest-frame) implied by the J$_{125}$ band photometric excess and magnification factor of the source.} \label{table:properties}
\begin{tabular}{lccccccccccc}
\hline
 Cluster & ID  & $\rm{z_{spec}}$ & RA & DEC  & I$\rm_{814}$  & UV slope ($\beta$) & EW$\rm_{[OIII]+H\beta}$ (\AA)& Magnification factor ($\mu$) \\ 
 \hline 
	ACT-CL0102-49151 & 4854 & 1.579 & 01:03:04.619	& -49:17:04.62 & 24.2 & -1.9  &  620$\pm$220~\AA & $1.41_{-0.12}^{+0.12}$ \\ 
	RXCJ0232.2-4420 & 588 & 1.645 & 02:32:16.124 & -44:20:55.72 & 23.6 & -2.3 & 1990$\pm$200~\AA & $7.96_{-1.72}^{+0.02}$  \\
 	PLCK G287.0+32.9 & 5648 & 1.720 & 11:50:52.800 & -28:06:03.24 & 24.4 &  -1.9 & 1120$\pm$150~\AA & $13.6_{-3.2}^{+0.4}$\\
	RXCJ0911.1+1746 & 612 &  1.727 & 09:11:09.912 & 17:46:54.84	& 25.2 & -2.2 & 1850$\pm$170~\AA & $3.8_{-0.1}^{+0.1}$ \\
 \hline
\end{tabular}
\label{table:observations}
\end{table*}

\subsection{Optical spectroscopy} 

\begin{table*}
\caption{Details of LDSS3 and IMACS spectroscopic observations.  From 
left to right, the columns denote the cluster name, cluster redshift, mask name, RA and DEC of mask center,  Instrument, 
date of observations, positional angle of masks and total exposure time per mask. Further details are provided in \S2} \label{table:observations}
\begin{tabular}{lccccccccccc}
\hline
 Cluster & Cluster  & Mask & \multicolumn{2}{c}{Mask Center}  & Instrument & Dates  &  PA & $\rm{t_{exp}}$ \\ 
   &redshift& & RA & DEC & &  & (deg) & (ks) \\ 
    
  \hline 
  
RXCJ0232.2-4420 & 0.2836 & rxc0232 & 02:32:17.683 & -44:20:35.67 & LDSS3 & 2016 Nov 27  & 196 & 7.2 \\
RXCJ0911.1+1746 & 0.5049 & rxc0911	&  09:10:56.692 & 17:49:06.94	& IMACS & 2019 Mar 6-7  & -30 & 20.4  \\
PLCK G287.0+32.9 & 0.3900 & p287 & 11:51:04.021 & -28:04:56.58 & IMACS & 2019  Mar 6-7  & 40 & 13.2\\
ACT-CL0102-49151 & 0.8700 & elgordo &  01:02:58.980 & -49:16:01.52 & LDSS3 & 2016 Aug 2  & 185 & 7.2 \\

\hline
\end{tabular}
\label{table:observations}
\end{table*}
The four EELGs described above were observed over three Magellan observing runs between 2016 and 2019. 
For the first two observing runs, we used the Low Dispersion Survey Spectrograph 3 (LDSS3) spectrograph on the
Magellan Clay telescope, targeting the cluster fields with ACTCL0102-4854 and RXCJ0232-588. These observing runs formed part of a larger spectroscopic survey of lensed galaxies in the RELICS fields
using University of Michigan and University of Arizona time allocation that we will present in a future paper (Mainali et al. 2019, in prep). 
In our most recent run, we used the Inamori-Magellan Areal Camera and Spectrograph (IMACS; \citealt{Dressler2011}) on the
Magellan Baade telescope to target the fields with RXCJ0911-612 and PLCKG287-5648.  Below we first describe the LDSS3
observations and reduction, then detail the same for IMACS.  Details of the observing setup are provided in Table 2.

We utilize LDSS3 in multi-object mode, designing two masks targeting the ACTCL0102 and RXCJ0232 fields. 
In addition to the EELGs discussed in \S2.1, we include other gravitationally-lensed systems and cluster members.
Slit widths were set to 1\farcs0, while
the typical slit lengths were 6\farcs0.  We used the VPH-ALL grism (400 lines mm$^{-1}$) with no order
blocking filter, providing continuous spectral coverage from 4000~\AA\ to 10000~\AA\ with a 
spectral resolution of 7.7~\AA. This setup allows us to detect CIII] throughout the redshift
range selected by our color selection.  For sources at $z<1.68$, we should also be able to
confirm redshifts by detection of the [OII] doublet.   We observed the ACT-CL0102 field on 2016 Aug 02
and the RXCJ0232 field on 2016 Nov 27.  Both masks were observed for 2 hours with typical
seeing of $0\farcs7$.

The LDSS3 data were reduced using the publicly available Carnegie
Observatories System for MultiObject Spectroscopy (COSMOS) pipeline.\footnote{http://code.obs.carnegiescience.edu/cosmos}
We obtained bias fields during each afternoon prior to observations. Flat fields were obtained using
quartz lamps, and arcs were observed using HeNeAr lamps. Both flat fields and arcs were taken
during the nights prior to observing each field.
The pipeline performs bias subtraction, flat-fielding, and a wavelength calibration
using the comparison arcs. The wavelength solutions have typical rms of 1.5~\AA. The pipeline does 
sky subtraction using the optimal method from \citet{Kelson2003} before outputting the final 2D spectra.
Finally, the 1D spectrum is produced using a boxcar extraction with an aperture of 1\farcs1 (6 pixels). 
The final 1D error spectrum is extracted from the output 2D error spectrum generated by the pipeline.
We observed a standard star to 
compute instrumental response across the detector. The absolute flux calibration is
then obtained from the known continuum magnitudes of several slit stars that are placed on our
masks.  The LDSS3 spectra provide a median line flux limit (3$\sigma$) of 1.3$\times$10$^{-17}$ erg cm$^{-2}$ s$^{-1}$
in the wavelength range  5000-6000 \AA. This flux limit provides rest-frame equivalent width limits (3$\sigma$) as low as
$\sim$3-5~\AA\ for CIII]  over the same wavelength range.

We observed the PLCK G287 and RXCJ0911 fields using IMACS in multi-object mode.  As with the
LDSS3 observations, we designed slitmasks that include a mixture of the EELGs, candidate lensed galaxies,
and cluster galaxy members.  For the PLCK G287 mask, we also placed a slit on one of the other images of EELG
PLCKG287-5648 (source 2.2 using the \citealt{Zitrin2017} nomenclature). 
We used the 300 lines/mm grating blazed at an angle of 17.5$^{\circ}$ with the f/2 camera. The grating is optimized to
provide wavelength coverage from 3900~\AA\ to 8000~\AA, covering CIII] and other UV metal lines in the expected
wavelength range.  We used a slit width of 1\farcs0 that provided a spectral resolution of 6.7~\AA.

Both masks were observed on 2019 March 6-7.  Conditions were clear throughout the observations
with an average seeing of 0\farcs7.  Because of the fainter continuum magnitudes of the EELGs in
these fields, we obtained longer exposure times ($\sim$3.5-5.5 hr) in order to
reach sufficient depth to detect emission lines with rest-frame equivalent widths as low as $\sim$3-5~\AA. 
The data were reduced using the same pipeline and procedure as described above. Similar to the LDSS3 spectra, the absolute calibration 
is performed using continuum magnitudes of slit stars placed on our masks.
We reached median line flux sensitivities (3$\sigma$) of 7.2$\times$10$^{-18}$ erg cm$^{-2}$ s$^{-1}$
(RXCJ0911-612) and 9.6$\times$10$^{-18}$ erg cm$^{-2}$ s$^{-1}$ (PLCK G287-5648)
in the wavelength range (5000-6000~\AA) where CIII] is situated, enabling constraints to be
placed on lines with rest-equivalent widths as low as $\sim$5~\AA.

\begin{figure*}
\centering
\includegraphics[scale=0.7]{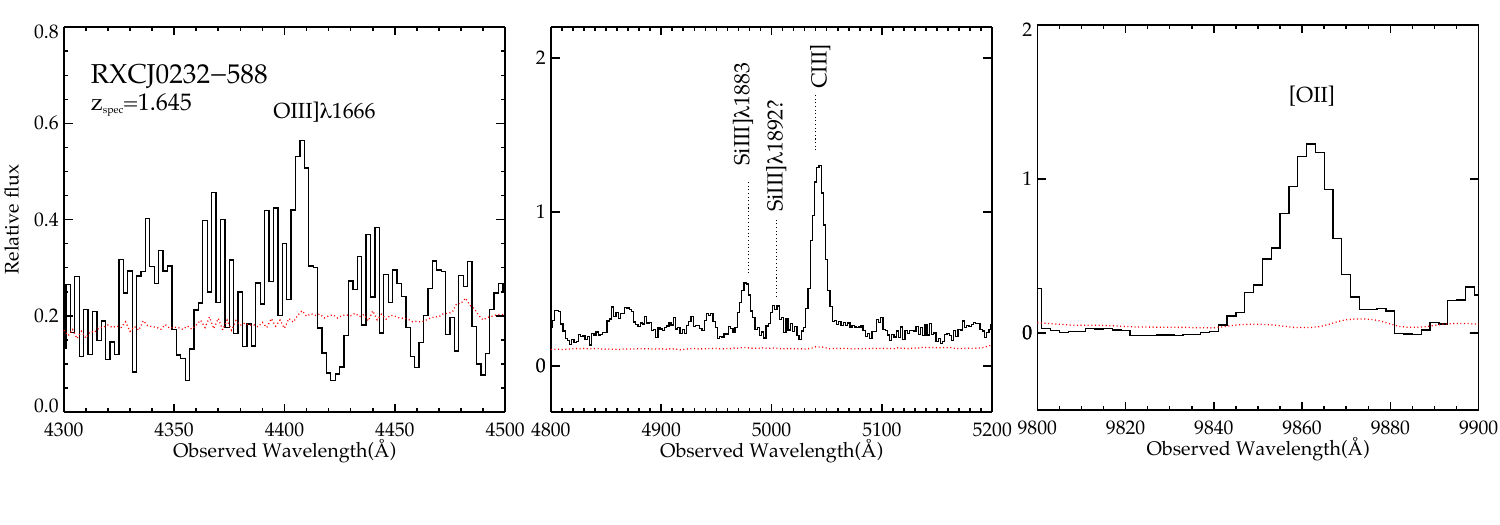}

\caption{ Magellan/LDSS3 spectrum of RXCJ0232-588 showing OIII]$\lambda$1666 line (left), SiIII]$\lambda$1883 and blended CIII]$\lambda\lambda$1907,1909 (middle) and blended [OII]$\lambda\lambda$3727,3729 (right). The black curve represents flux level and red dotted line represents 1-$\sigma$ error level in each panel. }
\end{figure*}

\subsection{Near Infrared Spectroscopy of RXCJ0232-588} 
We recently initiated  near-infrared spectroscopic follow-up of the EELGs, targeting 
RXCJ0232-588 with Folded-port InfraRed Echellette (FIRE; \citealt{Simcoe2013}) on the Magellan Baade telescope. 
Spectra in the near-infrared are required to detect  strong rest-optical nebular lines 
(i.e., [OII], H$\beta$, [OIII], H$\alpha$) which  constrain 
the physical conditions of the ionized gas, allowing us to better understand the 
properties most important for regulating the strength of the UV metal lines.  
Observations were obtained on 2018 September 02.  FIRE was operated in echelle
mode, providing spectral coverage between 0.82 and 2.51 $\mu$m. 
The observations were carried out using a slit width of 1\farcs0, resulting in a 
resolving power of R=3600. RXCJ0232-588 was observed for a total on-source integration 
time of 3 hours. Throughout the observations, the conditions  were excellent with an 
average seeing of 0\farcs5.  

The  spectrum of RXCJ0232-588 was reduced using standard routines in the FIREHOSE data reduction 
pipeline\footnote{wikis.mit.edu/confluence/display/FIRE/FIRE+Data+Reduction}. The pipeline uses 
lamp and sky flats for flat fielding. Two dimensional sky models are then iteratively calculated following \citet{Kelson2003}.  
The wavelength solutions are provided by fitting OH skylines in the spectra. Flux calibration and telluric corrections 
to the data are applied using  A0V star observations.  Finally, we the 1D spectrum was extracted using a boxcar 
with aperture of 0\farcs9 (15 pixels).

\section{Results}

Here we present the results from our spectroscopic observations of lensed EELGs.   In all cases, the spectra 
confirm the redshifts to lie in the range ($1.57<z<1.79$) predicted by the J$_{\rm{125}}$-band excess.
We first present redshift measurements and constraints on CIII] and other UV lines (\S3.1) before 
discussing the rest-optical lines detected in the FIRE spectrum of RXCJ0232-588 (\S3.2). 

\begin{figure}
\hspace{-0.3in}
\vspace{-0.3in}
\includegraphics[scale=0.4]{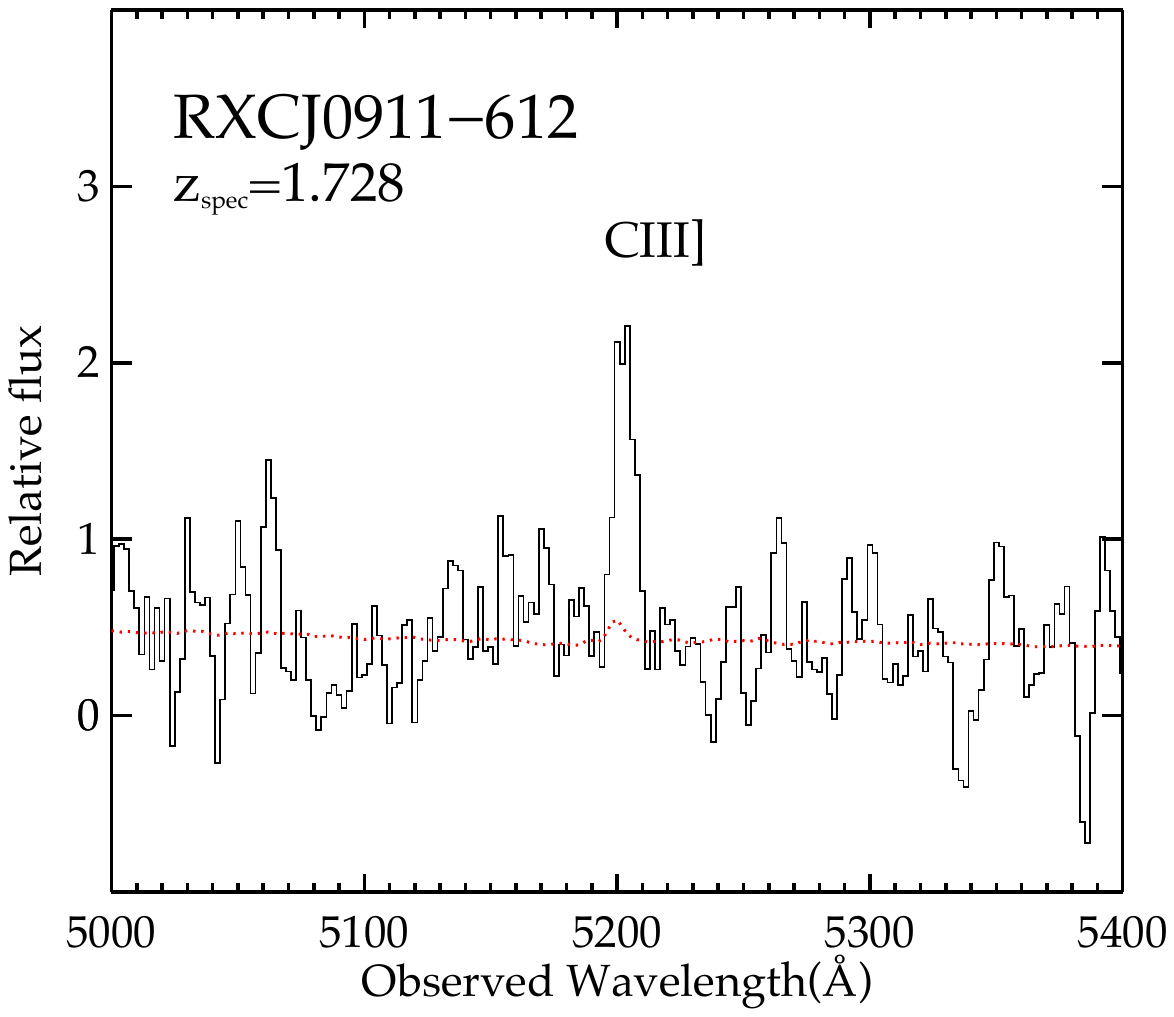}
\caption{ Magellan/IMACS spectrum of RXCJ0911-612 showing detection of blended CIII]$\lambda\lambda$1907,1909. The black curve in the plot represents flux level and red dotted curve denotes 1-$\sigma$ error level in the spectrum. }
\end{figure}

\subsection{Rest-UV Spectra: Probing CIII] Emission in  EELGs}   
\subsubsection{RXCJ0232-588 (EW$\rm_{[OIII]+H\beta}$=1990~\AA)}
The LDSS3 spectrum of RXCJ0232-588 covers rest-frame wavelengths between 1654(1523) and 3891(3584)~\AA\ 
assuming the extrema of the  redshift range predicted by the J-band excess, $1.57<z<1.79$.
  The spectrum shows four emission lines at 4406.6~\AA, 4980.1~\AA, 5046.3~\AA\ and 9861.2~\AA\ together with a continuum trace between 4200 and 9200~\AA\ (Fig.~3).   
We derive a redshift solution of $z=1.645$ with the lines corresponding to 
OIII]$\lambda$1666, Si III]$\lambda$1883, and the blended CIII]$\lambda$1908 and [OII]$\lambda$3728 doublets.  
We also identify a tentative detection (S/N=1.9) at 5004.1~\AA. This corresponds to the location of Si III]$\lambda$1892 component (Fig.~3). 
We do not see emission from OIII]$\lambda$1661.  The upper limit on the non-detection
implies OIII] doublet ratio of 1666:1661$>$1 (at 3$\sigma$).  This limit is 
fully consistent with the observed doublet  ratio in similar galaxies (e.g., \citealt{Erb2010, Stark2014,Berg2019}).  

We measure line fluxes of each emission feature by directly integrating fluxes after subtracting continuum near each line.
We then correct for the small aperture losses via methods 
described in previous papers (e.g., \citealt{Stark2014}).  Briefly, we convolve the {\it HST} image 
with the ground-based seeing and calculate the fraction of the object falling on the slit.  We then do the 
same for the slit star that was used to compute the absolute flux calibration.  The ratio of the 
object and star slit losses gives the factor by which we must correct our measured line fluxes, accounting 
for the spatial extent of the sources.  In the galaxies considered in this paper, the sizes are sufficiently 
compact that this correction factor is negligible.  In the case of RXCJ0232-588, we derive a very small 
correction factor of 1.05, consistent with its compact nature.  We finally compute the equivalent widths, 
dividing the emission line fluxes by the continuum flux level near the line of interest.  The continuum is derived from the spectrum when it is 
detected confidently (S/N$>$10) in a window of 200 \AA\ surrounding the emission line in question.  In cases where the continuum is 
not detected at the desired wavelength, we use the value implied by the best-fit population synthesis model 
(see \S4.2).  

The rest-UV line measurements of RXCJ0232-588  are presented in Table 3.  It is clear from the Table 
that this source is one of the most extreme UV line emitters known outside the reionization era.  
In particular, the spectrum clearly shows a strong CIII] emission feature with EW$_{\rm{CIII]}}$=21.7$\pm$2.8~\AA. 
This is larger than the 20~\AA\ threshold that has been suggested is a signpost for photoionization from AGN or 
super-solar C/O ratios \citep{Nakajima2018}, although we note that the EW uncertainty is such that this 
source could have a CIII] EW just below 20~\AA.   As we will show in \S5, 
RXCJ0232-588 is  consistent with an extension of the relationship between CIII] and [OIII] EW derived 
from a compilation of sources in the literature, with the largest CIII] strengths seen in galaxies with the 
most extreme optical line emission. 
The other detected UV lines are also very strong (EW$_{\rm{OIII]\lambda1666}}=3.8\pm1.3$~\AA\ and 
EW$_{\rm{Si III]\lambda1883}}=4.9\pm1.2$~\AA), with values among the largest seen 
in metal poor star forming galaxies (e.g., \citealt{Vanzella2016, Vanzella2017, Senchyna2017, Berg2018, Berg2019}).
 The blended [OII] doublet is detected with EW$_{\rm{[OII]}}=100\pm11$~\AA, 
similar to the strengths seen in other $z\simeq 2$ EELGs \citep{Tang2019}.  We will present the 
rest-optical spectrum of this source in \S3.2, providing a more detailed picture of the physical 
conditions of the nebular gas, a topic we will come back to in \S4.1.

\subsubsection{RXCJ0911-612 (EW$\rm_{[OIII]+H\beta}$=1850~\AA)}

\begin{figure*}
\centering
\includegraphics[scale=0.72]{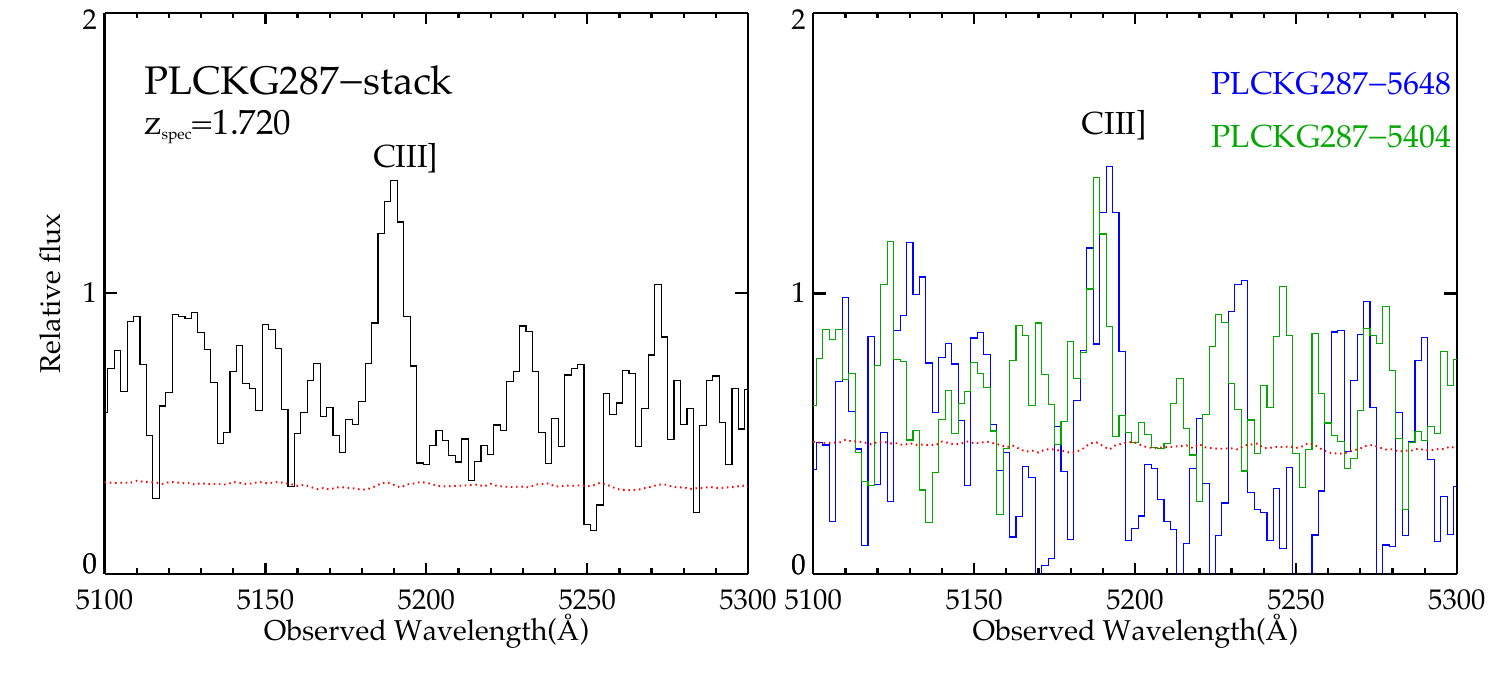}

\caption{ Magellan/IMACS spectra of PLCKG287-5648 and PLCKG287-5404, lensed images of the same galaxy. The left panel represents the stack of the two spectra where a blended CIII]$\lambda\lambda$1907,1909 line is visible. The right panel shows the spectrum of PLCKG287-5648 in blue and  PLCKG287-5404 in green. The CIII] emission feature is individually detected in each lensed images.}
\end{figure*}

The optical spectrum of RXCJ0911-612 spans 1556(1434)~\AA\ to 3502(3226)~\AA\ in the rest-frame, 
corresponding to the extrema of the  redshift range predicted by the J-band excess, $1.57<z<1.79$.  
 In this wavelength window, we expect CIII] to be our best 
probe of the redshift.  Over the redshift range predicted by the J$_{\rm{125}}$-band excess, 
CIII] is expected to be situated at observed wavelengths between 4903~\AA\ and 5322~\AA.  
We scan this wavelength window for lines and detect a confident (S/N=6.6) emission feature 
at 5204.4~\AA\ (Fig.~ 4).  We identify this as the blended CIII] doublet at a redshift of $z=1.728$.  
To calculate this redshift, we assume a rest-frame wavelength of 1907.709~\AA\ for the unresolved 
doublet (see \citealt{Erb2010}).  We detect no other confident (>5$\sigma$) emission 
features throughout the spectrum, consistent with expectations that CIII] should be the brightest line in this 
portion of the spectrum.  

We measure the flux and upper limits of emission lines using the same procedure described above.  The 
integrated flux of the combined CIII] doublet is 1.7$\pm$0.3$\times$10$^{-17}$ erg cm$^{-2}$ s$^{-1}$.   
The aperture correction relative to the slit star is again very small (1.08$\times$), not surprising given the 
compact size of RXCJ0911-612.  Using the continuum measured from the SED, we compute the 
rest-frame EW of CIII] and place upper limits on the other lines.  Similar to RXCJ0232-588, we find a value (EW$_{\rm{CIII]}}$=16.9$\pm$3.2~\AA) that is similar to what has been seen at $z>6$.  
Our measurements are presented in Table 3. 

\subsubsection{PLCKG287-5648 (EW$\rm_{[OIII]+H\beta}$=1120~\AA)}

The IMACS spectrum of PLCKG287-5648 covers  the same rest-wavelength window as for 
RXCJ0911-612.  We expect CIII] emission to be the strongest line in the spectrum. We search for line 
emission in the observed wavelength window where CIII] is expected to lie (4903~\AA\ to 5322~\AA).  
As shown in Fig.~ 5 (right panel, blue), we detect a 3.2$\sigma$ emission feature at an observed 
wavelength of 5189.8~\AA.  If this is indeed CIII], we should see emission at the same wavelength in 
PLCKG287-5404, another lensed image of the same galaxy (see discussion in \S2.1).  As is apparent in Fig.~5 (right panel, green), an emission line is seen at the same wavelength in the second image.  By median stacking the two spectra (Fig.~5, left panel), we detect the emission feature at higher significance (S/N=4.6). 
 We classify the detected feature as CIII] emission at a redshift of $z=1.720$, consistent with the redshift range implied by the strong J$_{\rm{125}}$-band excess.   We do not detect any other emission lines in the 
IMACS spectrum, as expected given the faint continuum of this EELG.  
 
Following the same procedures we have described above, we measure a flux of 9.1$\pm$2.0$\times$10$^{-18}$ 
erg cm$^{-2}$ s$^{-1}$ for the CIII] line in the stacked spectrum.  After making a small correction (1.21$\times$) 
for slit losses (relative to the slit star used for flux calibration), we compute the rest-frame EW. 
For this source, we use the continuum derived from the SED.  The resulting value (EW$_{\rm{CIII]}}$=4.8$\pm$1.1~\AA) 
is considerably lower than in the EELGs described above with more intense [OIII] emission.  Looking at the 
relationship between CIII] and [OIII]+H$\beta$ EW (see \S5), we see that  the measured CIII] strength in PLCKG287-5648 is lower than most sources with EW$\rm_{[OIII]+H\beta}\simeq 1000$~\AA).  Further
follow-up in the near-infrared should help clarify the origin of the scatter in this relationship.  
 
\subsubsection{ACTCL0102-4854 (EW$\rm_{[OIII]+H\beta}$=620~\AA)}

\begin{figure}
\centering
\includegraphics[scale=0.4]{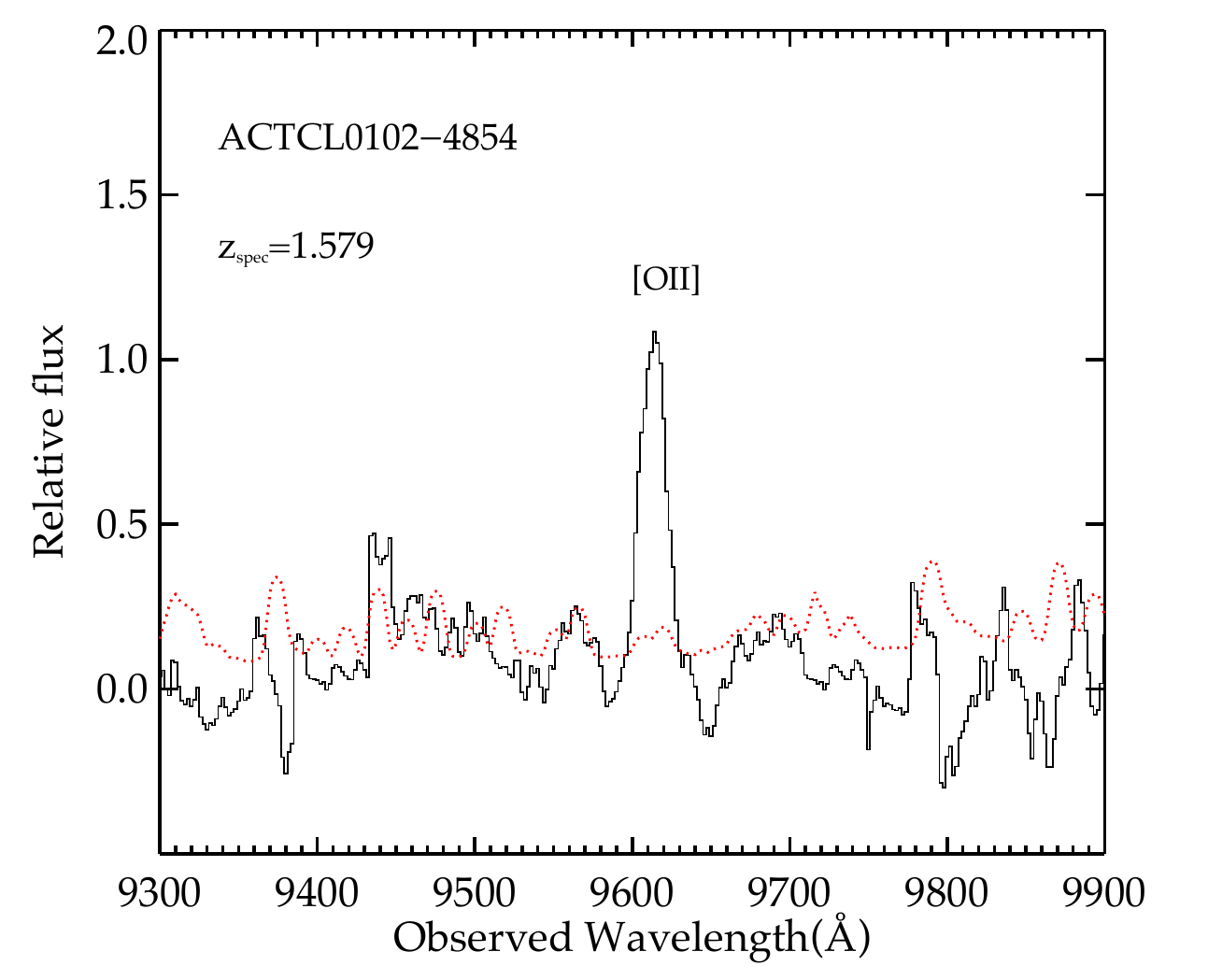}

\caption{ Magellan/LDSS3 spectrum of ACTCL0102-4854. The black curve and red dotted line represent flux and 1-$\sigma$ uncertainty, respectively. A blended [OII]$\lambda\lambda$3727,3729 is visible as a single prominent emission feature in the spectrum. }
\end{figure}

\begin{table*}
\begin{tabular}{lccccc}
\hline Object &  Line & $\lambda_{\rm{rest}}$  & $\lambda_{\rm{obs}}$  &  Flux & EW  \\
&  &  (\AA) &  (\AA)  &  ($\rm10^{-17}$ erg $\rm s^{-1}$ $\rm cm^{-2}$) & (\AA)  \\

\hline \hline
RXCJ0232-588 &     OIII]          & 1666.15 & 4406.6 & $2.7\pm0.9$ & $3.8\pm1.3$  \\
          		&	  Si III] & 1882.71 & 4980.1 & $2.6\pm0.6$ & $4.9\pm1.2$\\
		     &	       Si III] & 1892.03 & 5004.1 & $1.1\pm0.6$ & $2.1\pm1.2$\\
           		 &		CIII]  & 1907.71 & 5046.3 & $10.9\pm1.2$  & $21.7\pm2.8$ \\
           		& $\rm{[OII]}$ & 3728.6  &  9861.2 & $8.8\pm0.8$ & $99.8\pm11.2$  \\  
            \hline  
PLCKG287-5648 &     CIV       & 1549 & \ldots & <1.2 &  <6.6  \\
			&     He II          & 1640.42 & \ldots & <1.1 &  <5.5  \\
			&     OIII]          & 1666.15 & \ldots & <1.1 &  <5.5 \\
           		 &		CIII]  & 1907.71 & 5189.8 & $0.9\pm0.2$  & $4.8\pm1.2$ \\  
            \hline 

RXCJ0911-612 &     CIV       & 1549 & \ldots & <0.8 & <5.9  \\
			&     He II          & 1640.42 & \ldots & <0.7 & <7.3  \\
			&     OIII]          & 1666.15 & \ldots & <0.7 & <7.3  \\
           		 &		CIII]  & 1907.71 & 5204.4 & $1.7\pm0.3$  & $16.9\pm3.2$ \\ \hline
ACTCL0102-4854 & CIII]  & 1907.71 & \ldots & <1.1  & <4.0 \\   
		      & $\rm{[OII]}$ & 3728.6  &  9615.4 & $4.5\pm0.3$ & $59.2\pm5.7$  \\ 
            \hline                        
\end{tabular}
\caption{Magellan/LDSS3 and Magellan/IMACS emission line measurements of four EELGs presented in this paper. The upper limits are quoted at 3$\sigma$. }
\label{table:opticalspectra}
\end{table*}

The LDSS3 spectrum of ACTCL0102-4854 covers rest-frame wavelengths between 1654(1523) and 3891(3584)~\AA, assuming the extrema of the  redshift range predicted by the J-band excess, $1.57<z<1.79$.  
  We first scan for lines in the observed wavelength range where [OII] is 
expected (9583 to 10402~\AA).  A strong emission feature is readily apparent at 9615.4~\AA\ (Fig.~6).  This is consistent with a redshift of $z=1.579$, assuming a rest-wavelength of 3728.6~\AA\ for the unresolved doublet.   This rest-wavelength 
is calculated assuming a doublet flux ratio consistent with that seen in $z\simeq 2$ galaxies (e.g., \citealt{Sanders2016}).  
At this redshift, we would expect CIII] to appear at 4920~\AA.  No emission feature is seen in the vicinity of this wavelength, allowing us to 
place an upper limit on the line flux.  The other rest-UV lines (CIV, He II and OIII]) are all blueward of the LDSS3 
spectral coverage, so we cannot put constraints on their strength.

We derive the line flux of [OII] using the same methods as we described for the other three EELGs.  
The integrated line flux is 4.4$\pm$0.3$\times$10$^{-17}$ erg cm$^{-2}$ s$^{-1}$.  
We correct this flux by a very small factor (1.14$\times$), accounting for the excess slit losses of 
ACTCL0102-4854 relative to the stars used for flux calibration.  We finally calculate the rest-frame EW, 
using the continuum predicted near [OII] from the SED model.  The resulting value (EW$_{\rm{[OII]}}$=59.2$\pm$5.7~\AA) 
is within the range seen in similar strength [OIII] emitters \citep{Tang2019}.   The absence of 
CIII] suggests an upper limit of EW$_{\rm{[CIII]}}$=4.0~\AA, slightly weaker than average for 
galaxies with similar EW$\rm_{[OIII]+H\beta}$.   We present line measurements 
and upper limits in Table 3.  

\subsection{Rest-optical spectroscopy}

\begin{figure*}
\centering
\includegraphics[scale=0.7,angle=90]{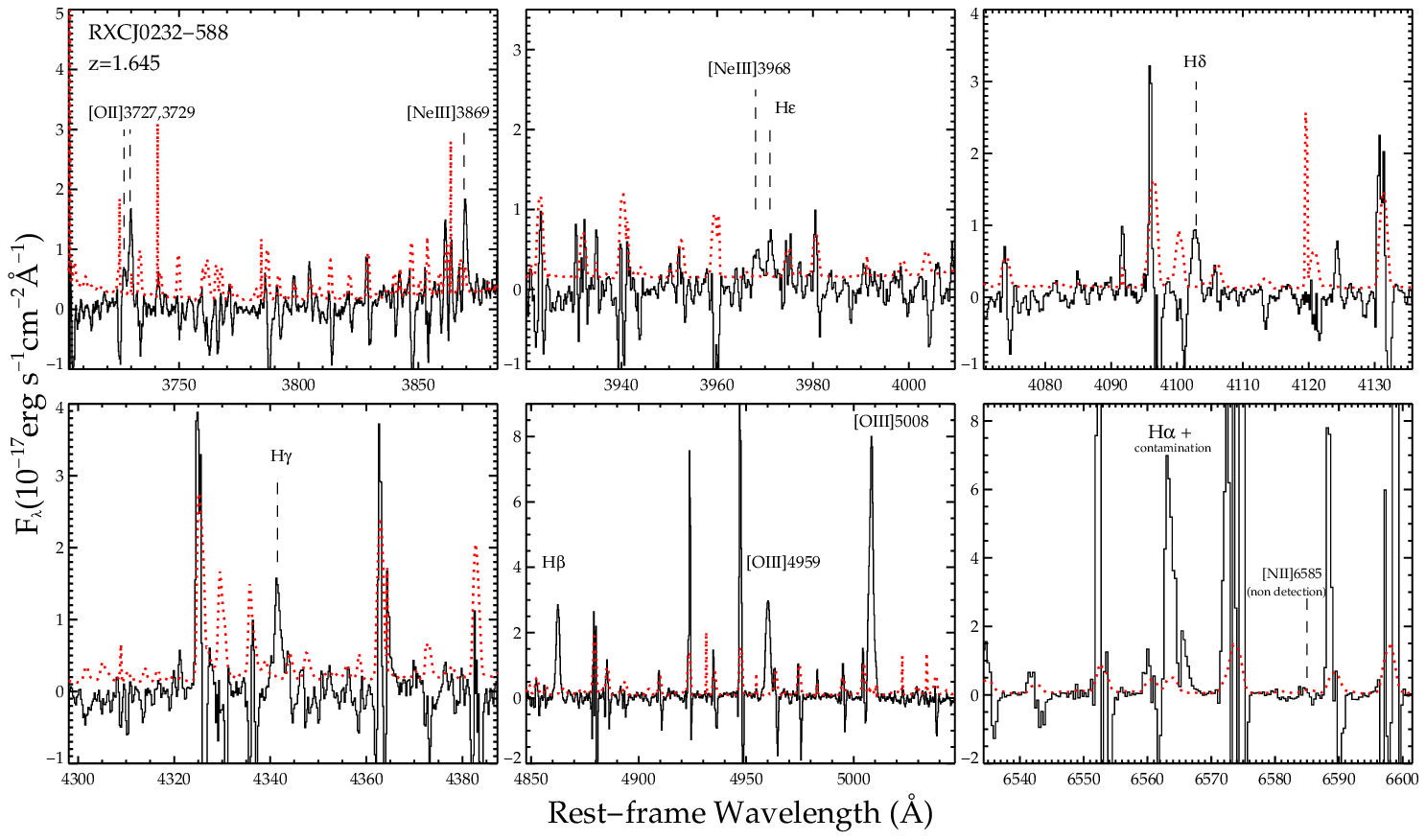}
\caption{FIRE spectrum of RXCJ0232-588 showing rest-optical emission lines. The black curve represents flux  and the red dotted line represents (1-$\rm{\sigma}$) uncertainty. 
The H$\alpha$ line is contaminated by a skyline. The spectrum reveals several optical emission features sensitive to ionization and abundance (see \S3.2). }
\label{fig:sample_spectra}
\end{figure*}

The FIRE spectrum of RXCJ0232-588 reveals ten unique emission features ([OII]$\lambda\lambda$3727,3729,$\rm[NeIII]\lambda3869$, 
$\rm[NeIII]\lambda3968$, H$\epsilon$,  H$\delta$, H$\gamma$,  H$\beta$, $\rm{ [OIII]\lambda4959,5007}$).  
Using the wavelength centroid calculated from the highest signal-to-noise nebular lines (H$\beta$, $\rm{ [OIII]\lambda4959,5007}$), we estimate a redshift of $z=1.6448$, consistent with the spectroscopic redshift 
derived from the LDSS3 data.  The line widths are narrow (FWHM=106.4 km s$^{-1}$, corrected for instrumental resolution). 
To compute line fluxes, we need to establish the absolute flux scale in the 
FIRE spectrum.  Since the [OII] emission line is also detected in our optical spectrum (see \S3.1), we can bootstrap the 
flux calibration of FIRE to that of LDSS3, the latter of which is determined with more confidence owing to the 
availability of slit stars observed simultaneously with the science spectra. This process takes into account flux calibration uncertainty in 
both FIRE and LDSS3 data whenever measuring line ratios that involve emission lines from the two spectra.

In Fig.~ 7, we present the 1D spectrum in the vicinity of the strong rest-frame optical emission lines.  
The majority of these lines are not significantly affected by skylines, enabling robust measurements 
of the integrated line flux.  We obtain these measurements using the MPFITPEAK routine in IDL, and 
corresponding errors are calculated using the error spectrum. 
Looking at the lower right panel in Fig.~7, it is clear that H$\alpha$ is 
coincident with a sky feature.  While the line is clearly detected, residuals from the sky subtraction 
make a reliable flux measurement challenging.  We will discuss this further below.  
Since the stellar continuum is undetected in the FIRE 
spectrum, we compute equivalent widths using the continuum derived from our best-fit broadband SED models 
(see Fig.~2 and \S4.2).   The measurement of line fluxes and corresponding equivalent widths are 
presented in Table 4. 

We characterize the impact of nebular attenuation using the observed flux ratios of the hydrogen 
Balmer lines.  Since H$\alpha$ is partially contaminated by a skyline, we consider the  ratio of 
H$\gamma$ and H$\beta$.   For case B recombination, no dust, and an electron temperature of 10$^4$ K, 
we expect to see F$_{\rm{H\gamma}}$/F$_{\rm{H\beta}}$ = 0.468 \citep{Osterbrock2006}.  The presence of 
dust will act to decrease this ratio relative to the theoretical value.   The observed flux ratio in the FIRE 
spectrum (F$_{\rm{H\gamma}}$/F$_{\rm{H\beta}}$=$0.48\pm0.06$) thus suggests little to no dust attenuation, consistent 
with the very blue UV continuum slope ($\beta=-2.3$) seen in the {\it HST} imaging (Fig.~2).  In the 
analysis that follows, we assume that the source is not significantly reddened  (E(B-V)$_{\rm{neb}}$=0.00), 
similar to other EELGs with similar [OIII] EW \citep{Tang2019}. In our analysis that follows in \S4, we 
thus take the observed line ratios as those intrinsic to the source.  Whenever required, we use the  
H$\alpha$ line strength implied by scaling the observed H$\beta$ flux by the theoretical flux ratio 
(F$_{\rm{H\alpha}}$/F$_{\rm{H\beta}}$=2.86) for case B recombination and 10$^4$ K gas \citep{Osterbrock2006}.  

The FIRE spectrum confirms the extreme emission lines implied by the photometry, revealing 
[OIII]$\lambda$5007 with EW=1426$\pm$57\AA\ and H$\beta$ with EW=325$\pm$23~\AA\ (Table 4).  
These values place RXCJ0232-588 among the most extreme optical line emitters known at high 
redshift (e.g., \citealt{Tang2019}), likely implying a very young stellar population (see \S4.2).  
The spectrum constrains numerous emission line ratios which are sensitive to the abundance 
and ionization state of the gas.  We list and define these in Table 5. 

 \begin{table}
\begin{tabular}{ccccc}
\hline  Line & $\lambda_{\rm{rest}}$ (\AA)& $\lambda_{\rm{obs}}$ (\AA)  &  $\rm F_{line}$/$\rm {F_{H\beta}}$ & EW (\AA)  \\

 \hline\hline            
 $\rm{[OII]}$  & 3727.13 & 9857.5 & $0.24\pm0.05$ & $59\pm10$\\
 $\rm{[OII]}$ & 3729.92 &9864.9 &$0.34\pm0.05$ &  $41\pm9$\\
   $\rm[NeIII]$ & 3869.66  & 10235.1 & $0.52\pm0.08$ & $91\pm7$ \\
     $\rm[NeIII]$ & 3968.2  & 10494.5 & $0.16\pm0.05$ & $28\pm9$ \\
       H$\epsilon$ & 3970.07 & 10851.3  & $0.16\pm 0.05$ & $28\pm9$ \\
  H$\delta$ & 4102.90& 10851.3  & $0.24\pm 0.03$ & $46\pm5$ \\
  H$\gamma$ & 4341.58& 11482.6 & $0.48\pm 0.05$ & $111\pm12$\\
 H$\beta$  &4862.55& 12860.5 & 1.00 & $325\pm23$ \\
 $\rm{ [OIII]}$ &4960.25 &  13118.9 & $1.34 \pm 0.03$ & $469\pm18$\\
  $\rm{[OIII]}$& 5008.27& 13245.9 & $4.08 \pm 0.03$ & $1426\pm57$\\
    $\rm{[NII]}$  & 6585 &\ldots  &$<0.08$ & $<77$\\
\hline 
\end{tabular}
F(H$\beta$)=(1.52$\pm$0.04)$\times\rm10^{-16}$ erg $\rm s^{-1}$ $\rm cm^{-2}$
\caption{Rest-optical emission line measurements of RXCJ0232-588. Emission line fluxes are presented relative to the H$\beta$.  The upper limits are 3$\sigma$. }
\label{table:optical_lines}
\end{table}

\begin{table*}
\begin{tabular}{lcc}

\hline  Quantity & Value & Notes  \\
 \hline \hline
 
 \multicolumn{3}{|c|}{Electron Density Sensitive Line Ratios} \\
 \hline 
        [OII] 3727/3729 & $0.75\pm0.15$ & $\rm n_{e}$=80$^{+160}_{-60}$ cm$^{-3}$ \\
       
 \hline 
 \multicolumn{3}{|c|}{Ionization Sensitive Line Ratios} \\
 \hline 
         O32 	     & 		 	$9.39\pm1.58$	& [OIII]4959,5008 /[OII]3727,3729 \\
         Ne3O2  &	 		$0.91\pm0.19$	& [NeIII]3869/[OII]3727,3729\\

\hline 
\multicolumn{3}{|c|}{Abundance-Sensitive Line Ratios} \\
 \hline 
 	N2 		&	$<0.03$	 &   [NII]6585/H$\alpha$ \\
         R23 	        &   	$6.01\pm0.13$	&  ([OIII]4959,5008+[OII]3727,3729)/H$\beta$  \\
 \hline     
 
 \multicolumn{3}{|c|}{Nebular Oxygen Abundances (Direct Method)} \\
 \hline 
        12+log(O/H) &	$7.60\pm0.24$		&	Using auroral line OIII]$\lambda$1666		 \\
 
 \hline          

\multicolumn{3}{|c|}{Nebular Oxygen Abundances (Strong Line Methods)} \\
 \hline 
        12+log(O/H)$\rm_{R23}$ &	$7.45\pm0.05$		&	Using \citet{Jones2015}		 \\
          12+log(O/H)$\rm_{Ne3O2}$ &  $7.84\pm0.05$			& 	Using \citet{Jones2015}		  \\  
 \hline    
      
\multicolumn{3}{|c|}{Inferred Gas Phase Abundance Ratios} \\
 \hline 
        log(C/O)  &	-$0.68\pm0.19$ &		\\
 \hline \hline                  
\end{tabular}
\caption{Measured and inferred properties of RXCJ0232-588. }
\end{table*}

\section{Analysis}

In this section, we explore the properties of the four gravitationally-lensed EELGs described in \S3.  We  
first explore the properties of RXCJ0232-588 in detail, leveraging the large number of emission lines 
that we have detected in our spectra of this source.  We then consider the properties of  our full sample, as 
implied by photoionization models.  In addition to constraining the gas conditions, these results reveal 
the population has  low stellar masses and young stellar populations, as would be expected for galaxies 
that have recently undergone a substantial upturn in their star formation. 

\subsection{Ionized Gas Properties of a Reionization Era Analog with Intense UV Metal Line Emission}

We use the emission line spectrum of RXCJ0232-588 to investigate 
the ionized gas properties of one of the few-known low mass star forming galaxies
that powers CIII] as strong as that found at $z>6$.  In addition to 
constraining the metallicity and ionization conditions of the galaxy, 
we seek to put this CIII] emitter in a broader context, comparing the observed 
line ratios to those of more massive and older star forming systems which are far more 
typical at $z\simeq 2$.  We use the insight gained from this analysis to comment on 
the likely powering mechanism of the CIII] emission. 

We first estimate the gas-phase oxygen abundance using the direct electron temperature (T$_{\rm{e}}$) method.  While 
we do not detect the [OIII]$\lambda$4363 auroral line, we do secure detection of OIII]$\lambda$1666, 
another auroral line commonly used to derive T$_{\rm{e}}$.  We use  PyNeb PYTHON package (version 1.1.8; \citealt{Luridiana2015}) 
to calculate T$_{\rm{e}}$([OIII]) from the observed flux ratio of OIII]$\lambda$1666 and 
[OIII]$\lambda$5007.  We fix the electron density to $n_{\rm{e}}$=80 cm$^{-3}$, as implied by the 
[OII] doublet ratio (Table 5).  Since the Balmer decrement suggests negligible nebular attenuation (\S3.2), 
we use the observed ratio of OIII]$\lambda$1666 and [OIII]$\lambda$5007 as intrinsic, propagating the 
errors on the Balmer decrement through to the temperature. 
With these assumptions, we derive an electron temperature of 16500$\pm$2400 K for the O$^{\rm{++}}$ zone. 
We next calculate T$_{\rm{e}}$([OII]) following the relation given in \citet{Perez-Montero2017a} which estimates T$_{e}$([OII]) using T$_{\rm{e}}$([OIII]) and an electron density ($n_{\rm{e}}$). 
This gives a temperature of T$_{e}$([OII])=14800$\pm$2100 K for the O$^{\rm{+}}$ zone.
Using these temperatures in $\rm{O^{+}}$ and $\rm{O^{++}}$ zones, we calculate $\rm{O^{+}/H^{+}}$ and $\rm{O^{2+}/H^{+}}$ 
from PyNeb. Combining the two ionic contributions, we calculate an oxygen abundance of 
12+log(O/H)=7.60$\pm$0.24 (0.08 Z$_{\odot}$; \citealt{Asplund2009}).  This is 3.5$\times$ lower than the 
direct method oxygen abundance inferred for $z\simeq2$ galaxies with more typical optical line EWs 
(e.g., \citealt{Steidel2016}).  We note that the oxygen abundances derived from collisionally excited lines (CELs) are 
known to be systematically lower than those derived from faint recombination 
lines (e.g., \citealt{Peimbert2002, Esteban2014}), with the latter thought to be a more 
reliable measure of the true gas-phase metallicity.  But regardless of the precise absolute 
value of O/H, the key point we wish to emphasize is that our data imply that RXCJ0232-588 has a significantly 
lower gas-phase metallicity than is found in the more massive high redshift star forming galaxies that have been studied in the MOSDEF and KBSS surveys.  

We now consider the ionization state of the gas in RXCJ0232-588.   This is  often parameterized as a 
dimensionless ionization parameter (U=n$_{\rm{\gamma}}$/n$_{\rm{H}}$), the ratio of the density of hydrogen ionizing photons 
that are incident on the gas and the number density of hydrogen atoms within the gas. The ionization parameter 
can be  constrained observationally by the  ratio of  emission lines from the same element with different ionization potentials. Most 
commonly used is the O32 index, defined as the flux ratio of the [OIII] and [OII] doublets.  The Ne3O2 index 
([Ne III]/[OII]) provides another useful constraint since the Ne/O abundance ratio does not vary 
substantially with O/H.  Over the past several years, the first statistical measures of O32 and Ne3O2 have been obtained at 
high redshift from the MOSDEF and KBSS surveys (e.g.,\citealt{Shapley2015, Sanders2016, Steidel2016, 
Strom2017}), revealing  evidence for a significantly higher ionization parameter than is common in local HII regions.  
The trend in redshift could reflect a combination of lower metallicities, harder ionizing spectra, or changes in the geometry 
of the nebular gas (see \citealt{Sanders2016} for a detailed discussion).  
However the EELGs we present in this paper are very different from the galaxies in these surveys, with lower stellar masses, large [OIII] EWs, lower metallicities, and a stellar population weighted much more toward very massive stars.  Recent work has demonstrated that both O32 and Ne3O2 increase with the [OIII] EW over $200~\AA< \rm{EW}_{\rm{[OIII]\lambda5007}}< 2000$~\AA\ \citep{Tang2019}, implying that the gas in the most extreme line emitters is much more highly ionized  than in the typical $z\simeq 2-3$ sources discussed above.  

Measurements of the ionization-sensitive ratios in RXCJ0232-588 support this picture, revealing 
nebular gas that is much more highly ionized than in typical $z\simeq 2-3$ galaxies.  The O32 ratio ($9.39\pm1.58$)  is 8$\times$ 
larger than the average of the MOSDEF galaxy sample \citep{Sanders2016} and 5$\times$ larger than is found in the  composite 
spectrum from the KBSS survey \citep{Steidel2016}. The Ne3O2 ratio ($0.91\pm0.19$) similarly points to 
highly-ionized gas, with a value that is 6$\times$ larger than is found in the KBSS composite \citep{Steidel2016}. 
The gas conditions in RXCJ0232-588 instead appear very similar to the EELGs presented in \citet{Tang2019}, with 
values of  O32 and Ne3O2 that are fully consistent with objects matched by [OIII] EW.   These results 
suggest a physical picture whereby the ISM in dwarf galaxies is found in a very highly ionized state for a short 
period following a burst of star formation. Given the frequent association between ionizing photon escape and 
large O32 (e.g., \citealt{Vanzella2016b,deBarros2016,Izotov2018, Fletcher2019}), it has been suggested that these bursts may initiate a  
short window where the ISM is conducive to substantial Lyman Continuum leakage in low mass systems 
\citep{Tang2019}. 

As explained in \S1, the C/O ratio of the nebular gas is another critical parameter for regulating CIII] 
emission line strengths.  While metal poor star forming galaxies are often found with sub-solar C/O ratios 
\citep{Berg2016, Berg2019}, it has been suggested that solar or super-solar C/O ratios are required to 
explain the large CIII] EWs that have been detected at $z>6$ if stars are responsible for powering the line 
emission (e.g., \citealt{Nakajima2018}). 
The LDSS spectrum of RXCJ0232-588 allows us to investigate whether this is the case for a 
system at $z\simeq 2$ with EW$_{\rm{CIII]}}$ similar to what has been seen in the reionization era.  
We used PyNeb to calculate the ratio of doubly ionized carbon and oxygen
from the flux ratios of CIII] and OIII]$\lambda$1666. To calculate the 
C/O ratio, we must apply an ionization correction factor (ICF) to $\rm{\frac{C^{+2}}{O^{+2}}}$
 \begin{eqnarray}
 \rm{ \frac{C}{O} = \frac{C^{+2}}{O^{+2}}}\times ICF.
 \end{eqnarray}
 
The ICF accounts for the possibility that volume fraction of C and O in their respective doubly-ionized states may not 
be identical.  \citet{Berg2019} have calculated the ICF as a function of the ionization parameter using photoionization 
models from CLOUDY v17.00 \citep{Ferland2013}.  They use BPASSv2.14 burst models \citep{Eldridge2016} 
with ages ranging between 10$^6$ to 10$^7$ years and stellar metallicities ranging from 0.05 to 0.40 Z$_\odot$.  The gas-phase metallicity 
is taken to be the same as the stars, and the ionization parameter is allowed to vary in the range $-3.0<\rm{log~U}<-1.0.$ 
To apply this to RXCJ02232-588, we must first calculate the ionization parameter of the nebular gas. To do so, we 
use the polynomial fitting functions for log U presented in \citet{Berg2019}. Taking the average value using the coefficients for the Z=0.05 and 0.10 Z$_\odot$ photoionization models and our measured value of O32, we find $\rm{log~U =-2.25\pm0.14}$, where the error bar corresponds to the scatter between the two metallicity models.  For this value of log U, we find an ICF of $1.02\pm0.04$ (see \citealt{Berg2019}), which 
implies $\rm log~C/O=-0.68\pm0.19$ for the total C/O ratio.  This is 0.4$\times$ the solar C/O ratio ($\rm log~C/O_\odot=-0.26$), consistent with the sub-solar values typically found in  metal poor systems in the literature \citep{Garnett1995,Erb2010,Christensen2012,Stark2014,Berg2016,Perez-Montero2017,Amorin2017,Senchyna2017,Berg2019}.  This suggests 
that it is possible to power the intense line emission seen in the reionization-era with sub-solar C/O ratios.  
 
The ionization and excitation conditions of the nebular gas can be further explored through investigation of the 
RXCJ0232-588 in the [OIII]$\lambda$5007/H$\beta$ versus [NII]$\lambda$6584/H$\alpha$ diagnostic 
diagram (the BPT diagram; \citealt{Baldwin1981}).  The flux ratios of RXCJ0232-588 place it in the upper left 
of the BPT diagram (Fig.~8), in a region consistent with the lines being powered by metal poor massive stellar 
populations.  The upper limit on N2 (log ([NII]$\lambda$6584/H$\alpha$) $<-1.56$) is indicative of low metallicity gas, as 
we have shown above.  The value of O3 ([OIII]$\lambda$5007/H$\beta$ = $4.08\pm0.03$) is very large, consistent with the 
gas being both highly ionized and elevated in T$_{\rm{e}}$.   While the O3 measurement is somewhat larger than the average 
value found in the MOSDEF survey (Fig.~8; \citealt{Sanders2016}), it is nearly identical to that derived from the  
composite spectrum of KBSS galaxies \citep{Steidel2016}.  While the larger ionization parameter of RXCJ0232-588 
shifts it toward larger O3, this effect is counterbalanced by the much lower O/H, resulting in an O3 measurement that 
is similar to that of the less extreme MOSDEF and KBSS galaxies.  
Recent investigations of Lyman-alpha emitters within the KBSS survey have also revealed many systems with low 
values of O3 \citep{Trainor2016}, likely also reflecting very low gas-phase oxygen abundances.   The position of 
RXCJ0232-588 in the O32 vs R23 diagram (Fig.~9) provides further information on the physical state of the gas. 
As we motivated above, the large O32 of RXCJ0232-588 is consistent with the trend between O32 and [OIII] EW 
presented in the EELG survey of \citet{Tang2019}.  But as can be seen in Fig.~9, RXCJ0232-588 has a lower value of R23 ($6.01\pm0.13$) than the majority of galaxies with similarly large O32. This is consistent with RXCJ0232-588 
having lower metallicity gas than the bulk of the EELGs from the CANDELS fields in \citet{Tang2019}, possibly a result of lensing allowing 
us to probe lower mass galaxies.  

One of the key goals in this paper is to ascertain the powering mechanism of the CIII] emission in RXCJ0232-588. 
The data presented here reveal that the ionized gas in the galaxy is metal poor, with an SED that is suggestive of a stellar population weighted toward extremely young stars (see \S4.2 for model constraints on the age, mass, and sSFR). 
These properties are very different from CIII] emitters with $\rm{EW>20}$~\AA\ presented in \citet{LeFevre2019}, with 
lower masses, much younger stellar populations, and no clear signatures of AGN activity.  
Given the low metallicity of the ionized gas in RXCJ0232-588, we expect the massive stars in the galaxy 
are likely to power a  hard EUV radiation field capable of powering strong nebular line emission.  We note that if the stellar metallicity (effectively set by the abundance of iron and iron peak elements)
is lower than implied by the gas-phase oxygen abundance due to a delay in iron production (e.g., \citealt{Steidel2016, Strom2018, Sanders2019}), this conclusion will only be strengthened.  Given the young age of the stellar population, we also expect a relatively weak underlying optical continuum. The combination of strong nebular emission and weak continuum provide a natural explanation for the extremely large CIII] EW seen in the LDSS spectrum of RXCJ0232-588.  

While massive stars are likely to provide a large source of EUV photons that contribute to the the CIII] 
intensity in  RXCJ0232-588, we currently cannot rule out the presence of other sources of ionization.  
There could be  contribution from a narrow-lined AGN (e.g., \citealt{Feltre2016, Volonteri2017}), provided it is low 
enough metallicity to not shift the  galaxy away from the upper left of the BPT diagram (e.g., \citealt{Groves2006, 
Izotov2008, Reines2013, Feltre2016}).  There also could be contribution from fast radiative shocks 
(e.g., \citealt{Allen2008, Jaskot2016}), but as with the AGN, we would require the gas to  be sufficiently low metallicity for consistency with the low [NII]/H$\alpha$ ratio 
\citep{Allen2008}.  Constraints on the strength of higher ionization lines in the far-UV (CIV$\lambda$1550, He II$\lambda$1640) should help disentangle whether shocks or AGN make a non-negligible contribution to the EUV 
output (e.g., \citealt{Feltre2016, Jaskot2016, Mainali2017}), but all current evidence appears  
consistent with massive stars dominating the ionizing output.  In the following sub-section, 
we will consider whether photoionization from a stellar population alone can self-consistently explain the 
CIII] emission and other spectral constraints in our EELG sample. Failure to do so might already point to the 
need to consider the alternative sources of ionization described above.  

\begin{figure}

\hspace{-0.3in}
\vspace{-0.3in}
\includegraphics[scale=0.4]{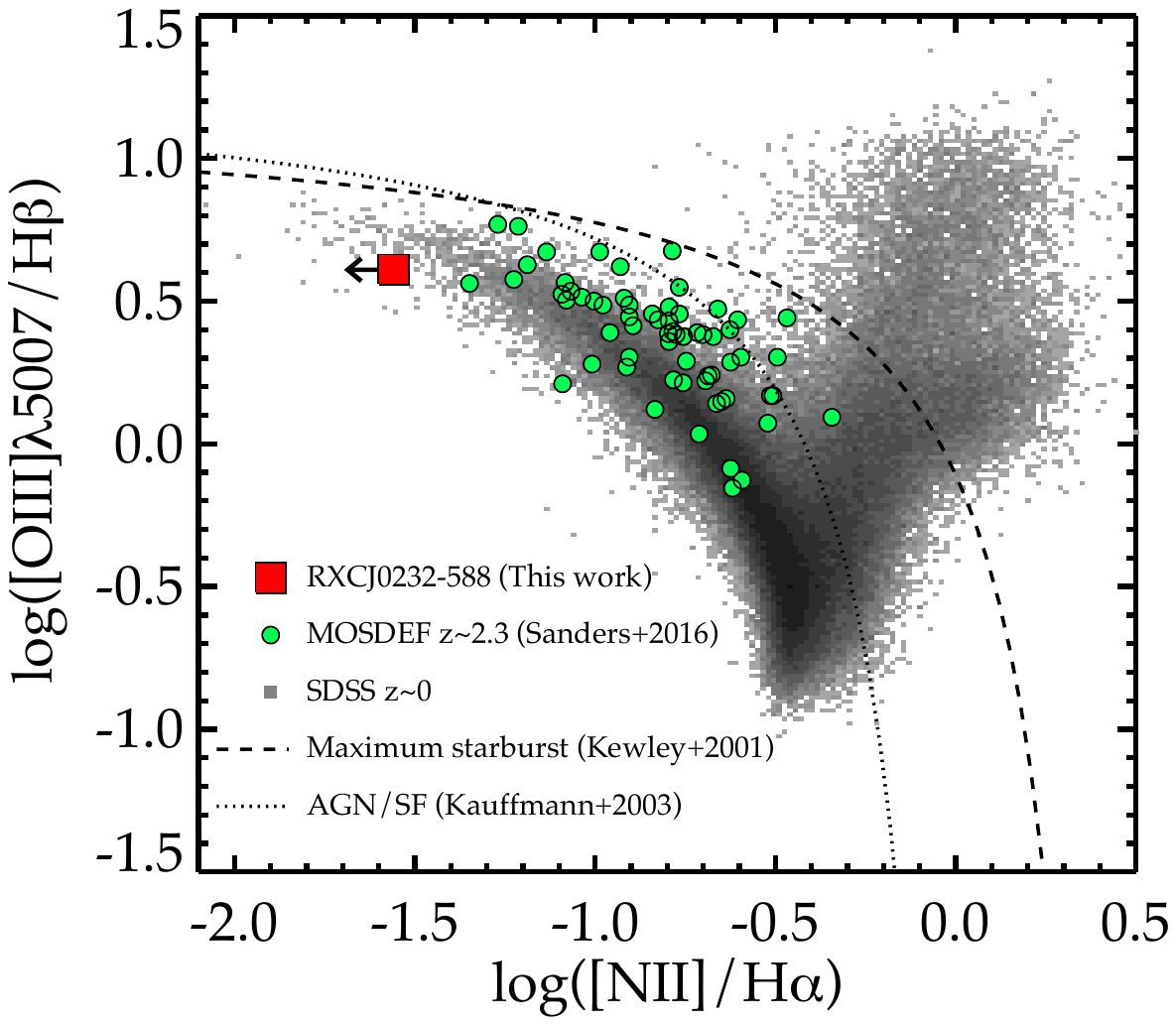}
\caption{ BPT diagram showing the position of RXCJ0232-588 (red square).  Grey data points represent $z\sim0$ SDSS sources. 
Green circles indicate $z\sim2.3$ galaxies from the MOSDEF survey. 
The dashed line depicts the maximum starburst model given by \citet{Kewley2001}, whereas the dotted line presents the AGN/SF demarcation from \citet{Kauffmann2003}. }
\end{figure}

\begin{figure}
\hspace{-0.3in}
\vspace{-0.3in}
\includegraphics[scale=0.4]{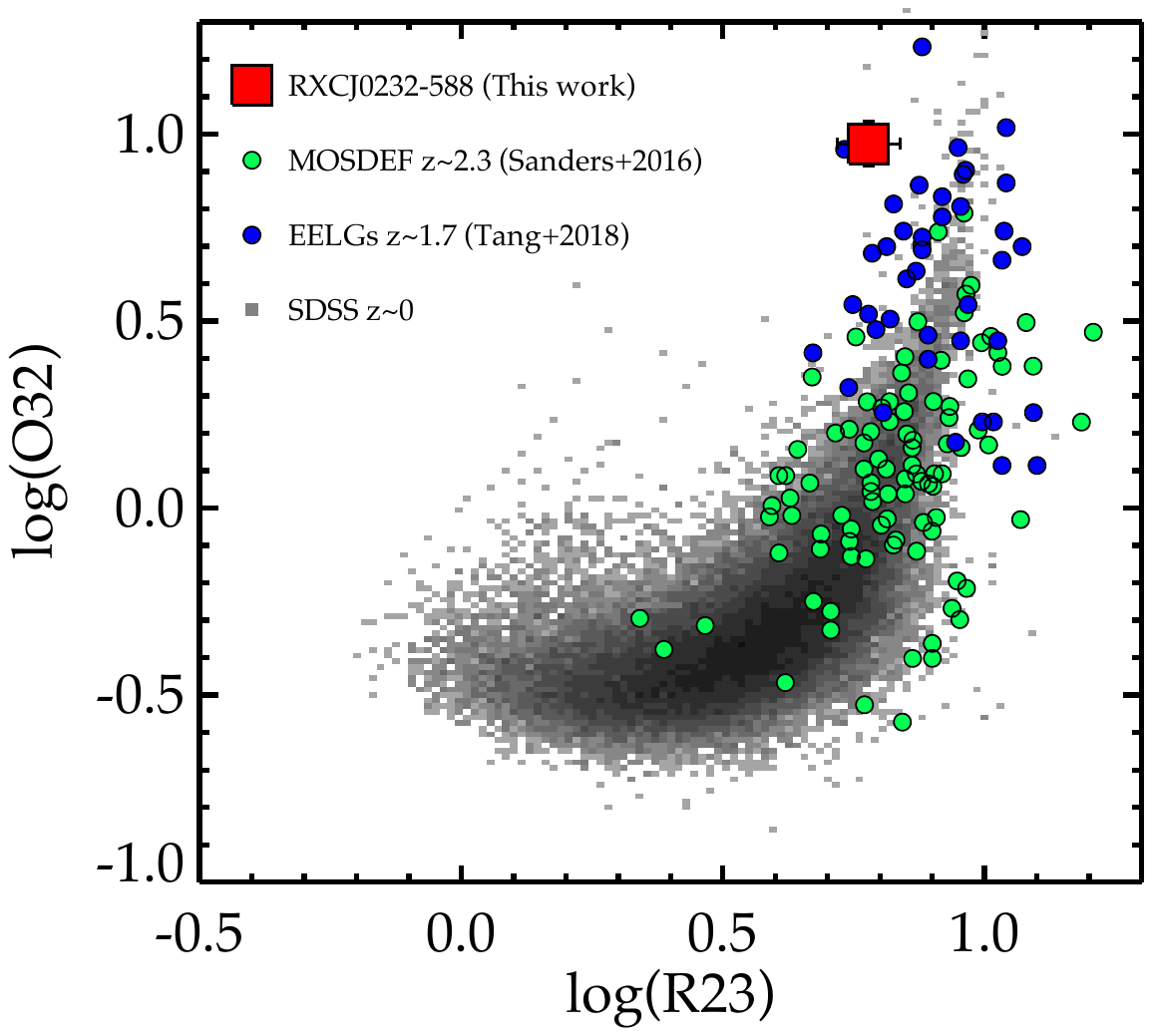}
\caption{Plot of O32 vs R23 demonstrating the position of RXCJ0232-588 (red square). Grey data points represent $z\sim0$ SDSS sources. Green circles indicate $z\sim2.3$ galaxies from the MOSDEF survey 
and blue circles represent EELGs from \citet{Tang2019}.}
\end{figure}

 \subsection{Photoionization modeling}
 
 \begin{table*}
\caption{Results from BEAGLE modeling of the four EELGs presented in this paper. From left to right, the columns 
give the object name, magnification-corrected stellar mass, magnification-corrected star formation rate,
 specific star formation rate, Ionization Parameter, Metallicity, Stellar Age, and V-band optical depth.} \label{table:photoionization}
\begin{tabular}{lccccccc}
\hline
 Object  & Stellar Mass  & SFR & sSFR & Ionization Parameter & Metallicity & Stellar Age & $\hat{\tau}_V$  \\ 
 	& log(M$_{\star}$/M$_{\odot}$)  &  (M$_{\odot}$yr$^{-1}$) & (Gyr$^{-1}$) & log(U$_S$) &  log(Z/$\rm{Z_{\odot}}$) & (Myr) & \\ \hline 
RXCJ0232-588 &$7.2_{-0.1}^{+0.1}$ &$4.7_{-1.0}^{+1.5}$ & $307.0_{-4.8}^{+5.1}$ & $-2.06_{-0.05}^{+0.05}$ & $-1.21_{-0.04}^{+0.02}$ & $3.3_{-0.1}^{+0.1}$ & $0.03_{-0.02}^{+0.02}$  \\
RXCJ0911-612 &  $7.1_{-0.1}^{+0.1}$&  $6.2_{-1.4}^{+1.2}$& $491.1_{-185.1}^{+291.7}$ & $-2.32_{-0.32}^{+0.38}$ & $-1.27_{-0.18}^{+0.14}$ & $2.0_{-0.7}^{+1.3}$ & $0.18_{-0.11}^{+0.12}$  \\
PLCKG287-5648 &$7.3_{-0.2}^{+0.2}$ & $1.2_{-0.5}^{+0.7}$ & $65.7_{-26.1}^{+48.7}$ & $-2.60_{-0.39}^{+0.29}$ & $-1.38_{-0.08}^{+0.18}$ & $15.2_{-6.5}^{+10.0}$ & $0.52_{-0.10}^{+0.09}$  \\
ACTCL0102-4854 & $7.9_{-0.1}^{+0.1}$ & $4.7_{-2.1}^{+1.2}$ & $58.0_{-13.9}^{+20.9}$ & $-2.45_{-0.13}^{+0.16}$ & $-1.39_{-0.08}^{+0.12}$ & $17.2_{-4.6}^{+5.5}$ & $0.31_{-0.06}^{+0.06}$ \\    
\hline
\end{tabular}
\label{table:modeling}
\end{table*}

We now investigate the physical properties of the four lensed EELGs in our sample, 
comparing the observed spectra and photometry to a suite of photoionization models. 
We make use of the Bayesian galaxy SED modeling and interpreting tool  BEAGLE  
(version 0.20.3; \citealt{Chevallard2016}).  BEAGLE is based on photoionization models of star forming galaxies in \citet{Gutkin2016} which combine the 
latest version of \citet{Bruzual2003} stellar population synthesis models with the photoionization code 
CLOUDY \citep{Ferland2013} to calculate emission from star and interstellar gas clouds.  Our goals in using 
BEAGLE are twofold.  First, we seek to infer bulk stellar population parameters (i.e., stellar mass, sSFR) implied by   
 the broadband SED and emission line properties.  Second, in cases where the spectra are sufficiently constraining,
we consider the ionized gas properties required by the data and explore whether stellar photoionization 
is capable of powering the observed CIII] emission.   

The models allow us to adjust the  interstellar  metallicity (Z$\rm_{ISM}$), the ionization parameter of the 
HII regions (U$\rm_{S}$), here defined at the edge of the Str\"{o}mgren sphere, and the dust-to-metal ratio ($\xi_{d}$), 
accounting for the depletion of metals on to dust grains.  We consider models with a hydrogen density of 
n$\rm_{H}$=100 $\rm cm^{-3}$.  We assume a C/O abundance that is 0.52$\times$ that of the solar value [(C/O)$_{\odot}\approx 0.44$], although we will also explore how models with solar C/O ratios would impact our findings.  
Our preferred use of a sub-solar C/O ratio is motivated by results demonstrating that galaxies with 
12+log O/H$<$8.0 tend to have C/O ratios that are less than 0.7$\times$ that of solar (e.g., \citealt{Berg2019}).   
We showed in \S4.1 that the UV spectrum of RXCJ0232-588 points toward sub-solar C/O ratios, providing 
further support for this assumption. We assume a constant star formation history, allowing the maximum stellar age to vary freely between 1 Myr and the age of the Universe at the redshift of the source we are considering. We assume \citet{Chabrier2003} initial mass function and a \citet{Calzetti2000} extinction curve.   We consider metallicities in the range of -2.2$\leq$log(Z/Z)$_{\odot}$$\leq$0.25.  Similar to our previous work, we assume that the interstellar metallicity is the same as the stellar metallicity (Z$_{\star}$=Z$_{\rm{ISM}}$).  The redshift of the models is fixed to the spectroscopically determined values for each source.  The  ionization parameter and dust-to-metal mass ratio are allowed to vary in the range -4.0$\leq$$U_{\rm S}$$\leq$-1.0 and $\xi_{d}$=0.1-0.5, respectively.   

We  fit the broadband photometry and the emission line equivalent widths simultaneously.  The 
broadband SED includes seven {\it HST} filters spanning the optical to near-infrared: B$_{\rm{435}}$, V$_{\rm{606}}$, I$_{\rm{814}}$, Y$_{\rm{105}}$, J$_{\rm{125}}$, JH$_{\rm{140}}$, H$_{\rm{160}}$. For one of the four EELGs (RXCJ0911-612), the F555W filter replaces  B$_{\rm{435}}$ and V$_{\rm{606}}$.  The BEAGLE fits to the photometry and emission-line EWs are overlaid on the SEDs in Fig.~2.  The model constraints on the magnification-corrected stellar masses, star formation rates, dust content ($\tau_V$), and sSFRs on the four EELGs are given in Table 6.  As 
can be seen in Fig.~ 2, the models reproduce the broadband SEDs.  They are also able to 
reproduce the rest-UV line measurements, and in the case of RXCJ0232-588, they match the 
rest-optical emission lines (see Table 7).  

The  stellar masses of the EELGs are very low, ranging from 1.3$\times$10$^7$ M$_\odot$ (RXCJ0911-612) to 
7.9$\times$10$^7$ M$_\odot$ (ACTCL0102-4854).   The fits suggest that there is little dust attenuation in 
these systems ($\tau_V= 0.03 - 0.52$), consistent with the blue UV continuum slopes that we observed in the 
broadband data (\S3.1).   As expected, we find that the sSFRs are very large and broadly increase with the amplitude 
of the J$_{\rm{125}}$-band excess, ranging from 58 Gyr$^{-1}$ (ACTCL0102-4854) and 66 Gyr$^{-1}$ (PLCKG287-5648) 
to 310 Gyr$^{-1}$ (RXCJ0232-588) and 490 Gyr $^{-1}$ (RXCJ0911-612).
These sSFRs imply very young mean stellar ages ($\lsim 20$ Myr) for the assumed constant star formation history. The young stellar age reflects that a recent 
burst dominates the observed SED and may not necessarily imply absence of faint older stellar populations from a past star formation activity. 
Under an assumption that the star formation proceeded with a recent burst  on top of an evolved stellar population,
 we found that the stellar age of oldest stars could be several hundred Myr. For instance, in case of  RXCJ0232-588 when we assume a recent burst within the last 5 Myr 
 along with the presence of evolved stars (represented by a delayed star formation history), we found that the maximum stellar age could be as high as 260 Myr.
  We note that the composite stellar population model increases our stellar 
 mass estimates by 0.5 dex, since this takes into account older generation of stars in the galaxy. However, the implied age of the recent burst is still low (1.6 Myr).
 It is only at these young ages that the ratio of O to A stars is large enough to reproduce the large EW optical line 
emission.   The ages range from 2.0 and 3.3 Myr for RXCJ0911-612 and RXCJ0232-588 to 15 Myr and 17 Myr for 
PLCKG287-5648 and ACTL0102-4854.  At ages of less than 3 Myr, the O star population will not have 
had time to equilibrate and will  be weighted more strongly to the hottest O stars, resulting in a harder 
EUV spectrum.  We thus expect the models 
that reproduce RXCJ0911-612 and RXCJ0232-588 will have harder ionizing spectra 
than those of the other EELGs in our sample.  
  
In the case of  RXCJ0232-588, the rest-UV and optical spectra have enough emission line 
detections to enable characterization of the ionized gas properties.  For our assumed sub-solar C/O ratio, 
the implied metallicity is very low (log Z/Z$_\odot= -1.21^{+0.02}_{-0.04}$), the dust attenuation 
is minimal ($\tau_V$=0.03$^{+0.02}_{-0.02}$), and the ionization parameter is large (log U$_{\rm{S}}$ 
= -2.06$^{+0.05}_{-0.05}$).  These parameters are capable of reproducing the optical line ratios and CIII] EW within  
observational uncertainties (Table 7), and we note that the metallicity implied by the photoionization modeling
($0.06\pm0.01$ Z$_\odot$)  is fully consistent with that implied by the T$_{\rm{e}}$ analysis in \S4.1 ($0.08\pm0.03$
Z$_\odot$). We also consider models with C/O ratios consistent with the solar value.  In this case, the derived
properties remain very similar, but the model CIII] EW (29~\AA) significantly exceeds the value determined
observationally, providing additional support for a sub-solar C/O ratio.  
We thus find that there are viable sets of parameters (i.e., metallicity, age, ionization parameter) that can 
reproduce the large CIII] EW of RXCJ0232-588 with stellar photoionization.  
The BEAGLE modeling procedure suggests that such strong  UV metal lines are a natural byproduct of the radiation field 
and gas conditions associated with an extremely young (3 Myr) and low metallicity (0.06 Z$_\odot$) stellar population, 
 as might expected to appear during a burst of star formation.   While we cannot rule out other sources of 
ionization, the mere detection of CIII] with EW$>$20~\AA\ should not necessarily imply the presence of AGN.  

For RXCJ0911-612, we do not have a rest-optical spectrum, 
but the detection of strong CIII] together with the pronounced J$_{\rm{125}}$-band excess  is already 
enough to place some constraints on the range of gas properties that can reproduce the data.  
When we adopt the sub-solar C/O models, we find that the data prefer
low metallicity (log Z/Z$_\odot$= -1.27$^{+0.14}_{-0.18}$), minimal dust attenuation ($\tau_V$=
0.18$^{+0.12}_{-0.11}$), and a large ionization parameter (log U$_{\rm{S}}$ = -2.32$^{+0.38}_{-0.32}$).  
Not surprisingly, these properties are very  similar to those derived for RXCJ0232-588, 
although the uncertainties in the gas properties of RXCJ0911-612 are considerably larger, owing to the absence 
of rest-optical line constraints.  Nonetheless, all evidence points toward the observations being consistent with 
another young system dominated by low metallicity massive stars. 
We are unable to usefully constrain the ionized gas properties for the two objects where the rest-UV 
lines are weak (or undetected) and rest-optical spectroscopy is not yet available (PLCKG287-5648 and 
ACTCL0102-4854).   While the broadband SED suggests that these systems are slightly older than the two systems 
described above (Table 6), it is not clear whether the metallicity or ionization parameter is significantly different.  
Near-infrared spectroscopy will be required to usefully constrain the gas conditions, providing a complete 
explanation as to why these systems power weaker CIII] emission.

\begin{table}
\begin{tabular}{lccc}
\hline
Object &  Quantity   & Data &  Model  \\ \hline
RXCJ0232-588 & CIII] EW (\AA)   & $21.7\pm2.8$ & 20.5  \\
	& O32  & $9.39\pm3.10$ & 11.85  \\
	& R23 & $6.01\pm$2.06 & 6.47 \\          
	& [OIII]$\lambda$5007/H$\beta$  & $4.09\pm0.56$ & 4.47 \\
	& [OIII]$\lambda$5007 EW (\AA) & $1426\pm$156 & 1692 \\ \hline
RXCJ0911-612 &   CIII] EW (\AA) & $16.9\pm3.2$ &  15.9 \\ \hline
PLCKG287-5648   & CIII] EW (\AA)   & $4.8\pm1.1$ & 4.9 \\ \hline
ACTCL0102-4854  & CIII] EW (\AA)  & $<4.0$ & 3.8 \\
  	& [OII] EW (\AA)  & $59.2\pm5.7$ & 54.2 \\  \hline
 
\end{tabular}
\caption{Observed and BEAGLE predicted emission line properties of the galaxies presented in this paper. The model values represent the posterior median of the marginal posterior distribution.}
\end{table}

\section{Discussion} \label{sec:discussion}
 
 \begin{figure}
\hspace{-0.3in}
\vspace{-0.3in}
\includegraphics[scale=0.5]{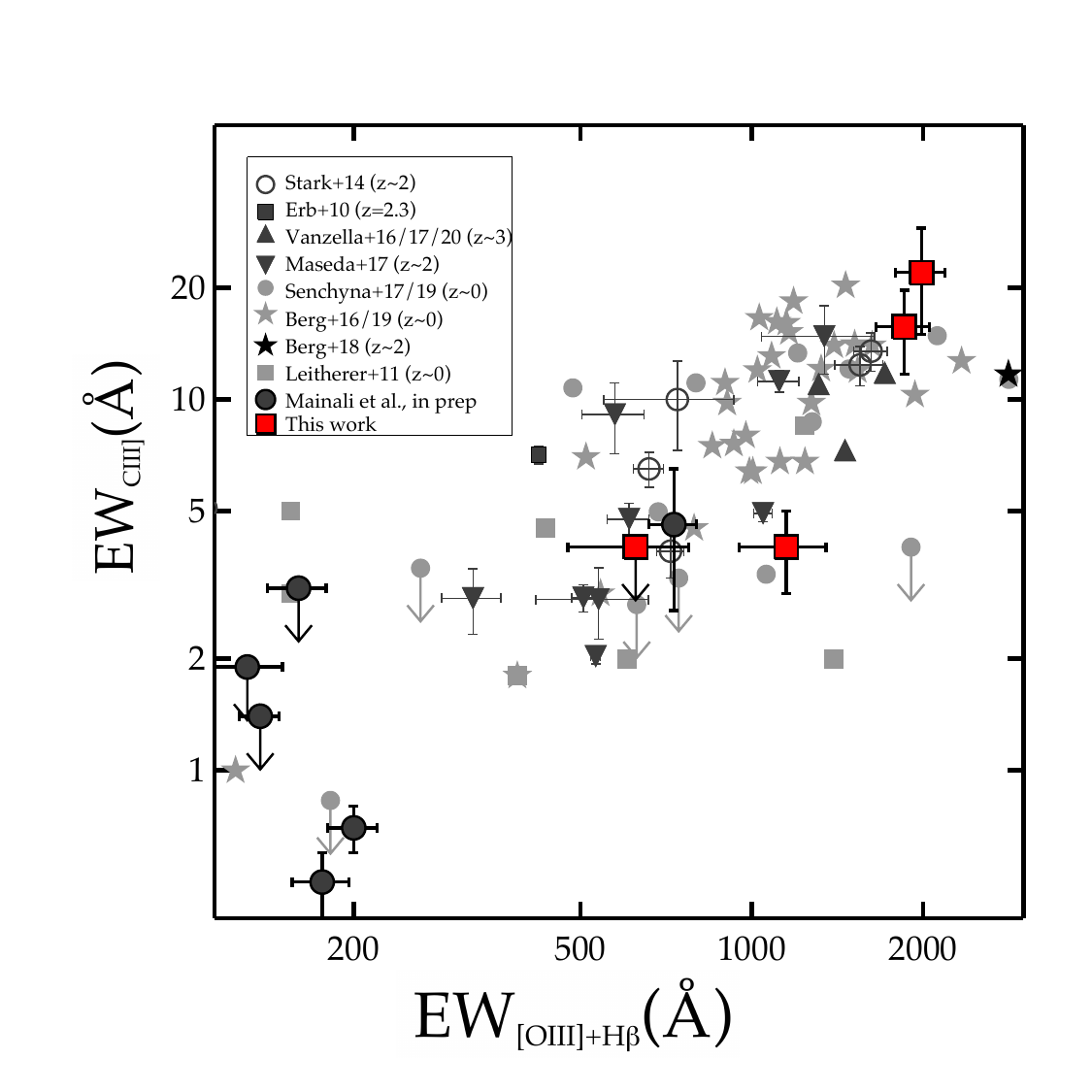}
\caption{Plot of CIII] rest frame equivalent width (EW$\rm_{CIII]}$) as a function of [OIII]+H$\beta$ rest frame equivalent width (EW$\rm_{[OIII]+H\beta}$). 
The red square represents data presented in this paper. The plot also includes a data compilation from the literature  \citep{Senchyna2017, Leitherer2011,Berg2016,Stark2014, Erb2010,Vanzella2016,Vanzella2017,Vanzella2020,Stark2017}. }
\label{fig:ciiivsoiii}
\end{figure}
 
The spectroscopic study of reionization-era galaxies is one of the primary science drivers behind the  
development of near-IR spectrographs for the current and future generation of O/IR telescopes.  
Whereas {\it JWST}  will be able to detect  strong rest-optical nebular lines such as [OIII] and H$\beta$ 
from $z>6$ galaxies,  ground-based O/IR facilities will be limited to the spectral features within the rest-UV.  
Identifying the optimal spectral features in the UV is particularly important for our ability to spectroscopically study  
the large sample of $z>6$ galaxies that will be identified photometrically by future imaging campaigns, for example, with the
 {\it Wide Field Infrared Survey Telescope} ({\it WFIRST}), as the areas covered will be much too wide for 
{\it JWST} follow-up. 
 
The discovery of intense UV metal line emission in $z>6$ galaxies suggests a path toward achieving this goal. 
While the faint lines are a challenge for existing facilities, their detection will become much easier with upcoming 
25-40m class telescopes, potentially opening the door for CIII] and CIV to be used for large redshift 
surveys in the reionization era. However it is worth emphasizing that we are currently very far from this goal.  
Each of the $z>6$ UV metal line detections are associated with relatively bright galaxies (H$<$26) for which the redshift was already known from 
Ly$\alpha$ emission; efforts to target CIII] in galaxies lacking spectroscopic confirmation have 
proven extremely challenging for current facilities (e.g., \citealt{Zitrin2015}). 
 Focusing on galaxies with Ly$\alpha$ redshifts is a natural starting place for the demonstration of method, but this approach almost certainly 
  leads to a biased spectroscopic sample, including only those sources in large enough ionized regions of the IGM 
for Ly$\alpha$ to be detectable at $z>7$.  

The last few years have seen several key steps taken toward the eventual use of CIII] and other 
faint UV lines as the primary means of spectroscopic confirmation at $z>6$.  Most importantly, 
we have improved our understanding of the stellar populations and gas conditions which support the strong line 
emission we are now seeing at $z>6$ (e.g.,\citealt{Stark2014, Rigby2015, Du2017, Senchyna2017}).  Intense metal line emission is found in galaxies that are both metal poor ($\lsim 0.4$ Z$_\odot$ for CIII]; 
\citealt{Senchyna2017}) and dominated by very young stellar populations. The latter trend is most clearly seen in the relation between CIII] EW and the [OIII] or H$\beta$ EWs. In Fig.~10, we present an updated compilation of the dependence of the CIII] EW on the [OIII]+H$\beta$ EW.  With the inclusion of the sources from our survey, it is now apparent that the CIII] EW reaches the values seen at $z>6$ (EW$_{\rm{CIII]}}>20$~\AA) in a subset of sources with [OIII]+H$\beta$ EW$ \simeq 2000$~\AA.  In the context of the photoionization models discussed in \S4.2, the CIII] detections 
at $z>6$ imply extremely young stellar populations expected within several Myr after a burst 
of star formation. 

This relationship guides the survey strategy for targeting CIII] 
 as the primary redshift indicator in metal poor star forming galaxies. The results of this paper demonstrate the 
feasibility of such efforts.  The sources we have targeted lacked spectroscopic confirmation 
but showed photometric flux excesses indicative of intense [OIII]+H$\beta$ emission at $z\simeq 1.6-1.8$.  
We were able to spectroscopically confirm the redshifts of the sources with the three largest broadband flux excess (i.e., the largest [OIII]+H$\beta$ EWs) through detection of CIII] emission.  These are among the first sources 
at high redshift for which CIII] was used as the primary means of spectroscopic confirmation. In each case, 
we guided our spectroscopic exposure times by expectations from the relationship between CIII] EW and 
[OIII]+H$\beta$ EW seen in Fig.~10.

The success of this approach motivates a renewed focus on current efforts to use CIII] as a redshift 
probe for reionization-era galaxy candidates.  The {\it Spitzer}/IRAC imaging necessary to identify intense rest-optical 
line emission is now available over large areas (e.g., \citealt{Strait2019,deBarros2019,Stefanon2019}), allowing efficient selection of bright galaxies 
likely to have detectable CIII] emission. In particular, Fig.~10 suggests that identification of sources 
with CIII] EW in excess of 10-20~\AA\ requires pre-selecting those galaxies with broadband flux excesses implying an 
[OIII]+H$\beta$ EW above 1500~\AA.  As discussed in \S1, this [OIII]+H$\beta$ EW is not typical of galaxies in the reionization-era, corresponding to roughly twice the value implied by composite SEDs \citep{Labbe2013}.
But by limiting spectroscopic searches to this subset of 
intense optical line emitters, it should be possible to efficiently identify CIII] emission independently 
of Ly$\alpha$ in deep near-infrared spectra.  In the era of the ELTs, these surveys can be extended 
to more typical sources with less extreme [OIII]+H$\beta$ EWs.  With spectroscopic flux limits expected 
from the ELTs (e.g., \citealt{Papovich2019}), the relationship in Fig.~10 suggests that it should be 
possible to identify CIII] in typical galaxies down to $\rm{H\simeq 26-27}$, thereby providing one of the only 
means of building large redshift samples at $z>6$ from future imaging surveys conducted by 
WFIRST and other wide area near-infrared missions, opening new doors for insight into early 
galaxies and reionization.  

\section{Summary}
The detection of intense CIII] and CIV emission in the spectra of $z>6$ galaxies has proven challenging to
interpret. Some have suggested that  AGN may be required to power the line emission (e.g., \citealt{Nakajima2018}) 
whereas other have suggested that the radiation field from young metal poor stellar populations is sufficient 
\citep{Stark2017}.  The difficulty stems in part from our poor understanding of the EUV radiation field 
powered by low metallicity stellar populations.  Without an improved reference sample of UV nebular line spectra of metal 
poor galaxies at lower redshifts, it will be impossible to distinguish between these two pictures.  

With the aim of beginning to assemble such a sample at $z\simeq 2$, we have obtained {\it Magellan}/IMACS and LDSS 
spectra targeting CIII]  in four gravitationally lensed reionization-era analogs identified in {\it HST} imaging of 
cluster fields from the RELICS survey.  The galaxies were selected to have a flux excess in the $\rm J_{125}$-band, indicative of strong [OIII]+H$\beta$ emission ($\rm{EW=500-2000}$~\AA), similar to what has been inferred 
from the SEDs of the $z>6$ galaxies with CIII] detections.  Our goal is to improve our understanding of the stellar populations and gas physical conditions that are required to power the intense CIII] emission that has been observed at $z>6$. We summarize our main findings below.

1. We detect CIII] emission in three out of the four EELGs in our sample, with the EW of CIII] scaling 
with the [OIII]+H$\beta$ EW.  The two objects  with the largest [OIII]+H$\beta$ EWs
are found to have CIII] strengths  approaching those seen at $z>6$ (EW$_{\rm{CIII]}}\simeq 17-22$~\AA).   
These results suggest that the relationship between CIII] EW and [OIII] EW continues well 
into the EELG regime, with the intense CIII] emission seen at $z\gsim 6$ linked to the 
most intense optical line emitters (EW$_{\rm{[OIII]+H\beta}}>1500-2000$~\AA).  

2.  Each of the four lensed galaxies in our sample are characterized by very low stellar masses 
(1.3-7.9$\times$10$^7$ M$_\odot$), large sSFRs (58-490 Gyr$^{-1}$) and low dust attenuation 
(V-band optical depth $\hat{\tau}_V$=0.0-0.5), 
consistent with a population of dwarf galaxies that have recently undergone a burst of star formation.   
The broadband SED and extreme optical line emission imply very young luminosity-weighted ages (for a 
constant star formation history), 
ranging from 2-3 Myr for the most extreme line emitters to 15-20 Myr for the two  EELGs with less 
prominent [OIII]+H$\beta$ emission. 
 
3. We have obtained a {\it Magellan}/FIRE near-infrared spectrum of RXCJ0232-588, providing insight 
into the gas conditions in  
one of the only low mass galaxies known at $z\simeq 0-2$ with an integrated CIII] EW as large as 
at $z>6$ (EW$_{\rm{CIII]}} >$20~\AA).  We find that the galaxy is very metal poor 
(12+log O/H =7.6$\pm$0.2) with a highly ionized ISM (O32=9.39) and a sub-solar C/O ratio 
($\rm{log~C/O=-0.68}$).  The properties of the source are very different from previously-known CIII] emitters 
with EW$>20$~\AA\ at $z\simeq 2-3$ \citep{LeFevre2019}, each of which have shown 
super-solar C/O ratios or evidence for AGN activity.   Instead these properties appear to be very similar to recent studies of metal poor
 low mass galaxies at similar redshifts \citep{Amorin2017,Vanzella2016,Vanzella2017, Berg2018}.

4. We fit the emission lines and broadband SEDs of  the EELGs in our sample using the BEAGLE 
photoionization modeling tool \citep{Chevallard2016}. We find that stars are capable of powering 
the observed CIII] emission without contribution from additional sources of ionization (AGN, shocks), in spite 
of the sub-solar C/O ratio.  In 
particular, the EUV radiation field associated with a young (2-3 Myr), metal poor stellar population 
(0.05-0.06 Z$_\odot$) is able to simultaneously reproduce the  CIII] emission,  optical nebular line ratios, and  broadband SED in our sample of extreme line emitters.  In this context, the increased detection rate of intense CIII] 
emission at $z>6$ \citep{Mainali2018} may suggest that such young, metal poor stellar populations are becoming more common in the reionization era.

5.  The three UV line emitters in our sample are among the first sources for which CIII] was successfully used 
as the primary means of redshift confirmation.  We discuss implications for the use of CIII] as a 
spectroscopic tool in the reionization era, arguing that the relationship between CIII] EW and [OIII] EW should already 
make it feasible to confirm redshifts of $z>6$ carefully-selected galaxies that have yet to be confirmed via Ly$\alpha$.  
In particular, $z\simeq 7-8$ dropouts with {\it Spitzer}/IRAC flux excesses indicative of EW$_{\rm{[OIII]+H\beta}}>1500-2000$~\AA\ are likely to have prominent CIII] emission, making spectroscopic confirmation tractable 
in sufficiently bright galaxies.  In the era of the ELTs, it should be feasible to detect CIII] in 
less extreme sources, providing one of the most efficient means of redshift confirmation for photometric 
sources identified in {\it WFIRST} imaging. 

\section*{Acknowledgements}

We thank Xinnan Du, Anna Feltre, Taylor Hutchison, Alice Shapley, and Casey Papovich for helpful discussions. 
DPS acknowledges support from the National Science Foundation  through the grant AST-1410155. 
This paper includes data gathered with the 6.5 meter Magellan Telescopes located at Las Campanas Observatory, Chile. 
This work is based on observations taken by the RELICS Treasury Program(GO-14096) with the NASA/ESAHST.
Program GO-14096 is supported by NASA through a grant from the Space Telescope Science Institute, which is 
operated by the Association of Universities for Research inAstronomy, Inc., under NASA contract NAS5-26555. 
The paper uses high level science products (HLSP) from the RELICS program, including catalogs and lens models,
 which were retrieved from the Mikulski Archive for Space Telescopes (MAST).

\bibliographystyle{mnras}
\bibliography{references}

\begin{thebibliography}{}
\makeatletter
\relax
\def\mn@urlcharsother{\let\do\@makeother \do\$\do\&\do\#\do\^\do\_\do\%\do\~}
\def\mn@doi{\begingroup\mn@urlcharsother \@ifnextchar [ {\mn@doi@}
  {\mn@doi@[]}}
\def\mn@doi@[#1]#2{\def\@tempa{#1}\ifx\@tempa\@empty \href
  {http://dx.doi.org/#2} {doi:#2}\else \href {http://dx.doi.org/#2} {#1}\fi
  \endgroup}
\def\mn@eprint#1#2{\mn@eprint@#1:#2::\@nil}
\def\mn@eprint@arXiv#1{\href {http://arxiv.org/abs/#1} {{\tt arXiv:#1}}}
\def\mn@eprint@dblp#1{\href {http://dblp.uni-trier.de/rec/bibtex/#1.xml}
  {dblp:#1}}
\def\mn@eprint@#1:#2:#3:#4\@nil{\def\@tempa {#1}\def\@tempb {#2}\def\@tempc
  {#3}\ifx \@tempc \@empty \let \@tempc \@tempb \let \@tempb \@tempa \fi \ifx
  \@tempb \@empty \def\@tempb {arXiv}\fi \@ifundefined
  {mn@eprint@\@tempb}{\@tempb:\@tempc}{\expandafter \expandafter \csname
  mn@eprint@\@tempb\endcsname \expandafter{\@tempc}}}

\bibitem[\protect\citeauthoryear{{Acebron} et~al.,}{{Acebron}
  et~al.}{2018}]{Acebron2018}
{Acebron} A.,  et~al., 2018, \mn@doi [\apj] {10.3847/1538-4357/aabe29}, \href
  {http://adsabs.harvard.edu/abs/2018ApJ...858...42A} {858, 42}

\bibitem[\protect\citeauthoryear{{Alavi} et~al.,}{{Alavi}
  et~al.}{2016}]{Alavi2016}
{Alavi} A.,  et~al., 2016, \mn@doi [\apj] {10.3847/0004-637X/832/1/56}, \href
  {http://adsabs.harvard.edu/abs/2016ApJ...832...56A} {832, 56}

\bibitem[\protect\citeauthoryear{{Allen}, {Groves}, {Dopita}, {Sutherland}  \&
  {Kewley}}{{Allen} et~al.}{2008}]{Allen2008}
{Allen} M.~G.,  {Groves} B.~A.,  {Dopita} M.~A.,  {Sutherland} R.~S.,
  {Kewley} L.~J.,  2008, \mn@doi [\apjs] {10.1086/589652}, \href
  {http://adsabs.harvard.edu/abs/2008ApJS..178...20A} {178, 20}

\bibitem[\protect\citeauthoryear{{Amor{\'\i}n} et~al.,}{{Amor{\'\i}n}
  et~al.}{2014}]{Amorin2014}
{Amor{\'\i}n} R.,  et~al., 2014, \mn@doi [\aap] {10.1051/0004-6361/201423816},
  \href {https://ui.adsabs.harvard.edu/abs/2014A&A...568L...8A} {568, L8}

\bibitem[\protect\citeauthoryear{{Amor{\'{\i}}n} et~al.,}{{Amor{\'{\i}}n}
  et~al.}{2017}]{Amorin2017}
{Amor{\'{\i}}n} R.,  et~al., 2017, \mn@doi [Nature Astronomy]
  {10.1038/s41550-017-0052}, \href
  {http://adsabs.harvard.edu/abs/2017NatAs...1E..52A} {1, 0052}

\bibitem[\protect\citeauthoryear{{Asplund}, {Grevesse}, {Sauval}  \&
  {Scott}}{{Asplund} et~al.}{2009}]{Asplund2009}
{Asplund} M.,  {Grevesse} N.,  {Sauval} A.~J.,   {Scott} P.,  2009, \mn@doi
  [\araa] {10.1146/annurev.astro.46.060407.145222}, \href
  {http://adsabs.harvard.edu/abs/2009ARA%26A..47..481A} {47, 481}

\bibitem[\protect\citeauthoryear{{Atek} et~al.,}{{Atek}
  et~al.}{2011}]{Atek2011}
{Atek} H.,  et~al., 2011, \mn@doi [\apj] {10.1088/0004-637X/743/2/121}, \href
  {http://adsabs.harvard.edu/abs/2011ApJ...743..121A} {743, 121}

\bibitem[\protect\citeauthoryear{{Baldwin}, {Phillips}  \&
  {Terlevich}}{{Baldwin} et~al.}{1981}]{Baldwin1981}
{Baldwin} J.~A.,  {Phillips} M.~M.,   {Terlevich} R.,  1981, \mn@doi [\pasp]
  {10.1086/130766}, \href {http://adsabs.harvard.edu/abs/1981PASP...93....5B}
  {93, 5}

\bibitem[\protect\citeauthoryear{{Berg}, {Skillman}, {Henry}, {Erb}  \&
  {Carigi}}{{Berg} et~al.}{2016}]{Berg2016}
{Berg} D.~A.,  {Skillman} E.~D.,  {Henry} R.~B.~C.,  {Erb} D.~K.,   {Carigi}
  L.,  2016, \mn@doi [\apj] {10.3847/0004-637X/827/2/126}, \href
  {http://adsabs.harvard.edu/abs/2016ApJ...827..126B} {827, 126}

\bibitem[\protect\citeauthoryear{{Berg}, {Erb}, {Auger}, {Pettini}  \&
  {Brammer}}{{Berg} et~al.}{2018}]{Berg2018}
{Berg} D.~A.,  {Erb} D.~K.,  {Auger} M.~W.,  {Pettini} M.,   {Brammer} G.~B.,
  2018, preprint, \href {http://adsabs.harvard.edu/abs/2018arXiv180302340B} {}
  (\mn@eprint {arXiv} {1803.02340})

\bibitem[\protect\citeauthoryear{{Berg}, {Erb}, {Henry}, {Skillman}  \&
  {McQuinn}}{{Berg} et~al.}{2019}]{Berg2019}
{Berg} D.~A.,  {Erb} D.~K.,  {Henry} R. B.~C.,  {Skillman} E.~D.,   {McQuinn}
  K. B.~W.,  2019, \mn@doi [\apj] {10.3847/1538-4357/ab020a}, \href
  {https://ui.adsabs.harvard.edu/abs/2019ApJ...874...93B} {874, 93}

\bibitem[\protect\citeauthoryear{{Bouwens}, {Illingworth}, {Oesch}, {Caruana},
  {Holwerda}, {Smit}  \& {Wilkins}}{{Bouwens} et~al.}{2015a}]{Bouwens2015b}
{Bouwens} R.~J.,  {Illingworth} G.~D.,  {Oesch} P.~A.,  {Caruana} J.,
  {Holwerda} B.,  {Smit} R.,   {Wilkins} S.,  2015a, preprint, \href
  {http://adsabs.harvard.edu/abs/2015arXiv150308228B} {} (\mn@eprint {arXiv}
  {1503.08228})

\bibitem[\protect\citeauthoryear{{Bouwens} et~al.,}{{Bouwens}
  et~al.}{2015b}]{Bouwens2015}
{Bouwens} R.~J.,  et~al., 2015b, \mn@doi [\apj] {10.1088/0004-637X/803/1/34},
  \href {http://adsabs.harvard.edu/abs/2015ApJ...803...34B} {803, 34}

\bibitem[\protect\citeauthoryear{{Bradley} et~al.,}{{Bradley}
  et~al.}{2014}]{Bradley2014}
{Bradley} L.~D.,  et~al., 2014, \mn@doi [\apj] {10.1088/0004-637X/792/1/76},
  \href {http://adsabs.harvard.edu/abs/2014ApJ...792...76B} {792, 76}

\bibitem[\protect\citeauthoryear{{Bruzual} \& {Charlot}}{{Bruzual} \&
  {Charlot}}{2003}]{Bruzual2003}
{Bruzual} G.,  {Charlot} S.,  2003, \mn@doi [\mnras]
  {10.1046/j.1365-8711.2003.06897.x}, \href
  {http://cdsads.u-strasbg.fr/abs/2003MNRAS.344.1000B} {344, 1000}

\bibitem[\protect\citeauthoryear{{Byler}, {Dalcanton}, {Conroy}, {Johnson},
  {Levesque}  \& {Berg}}{{Byler} et~al.}{2018}]{Byler2018}
{Byler} N.,  {Dalcanton} J.,  {Conroy} C.,  {Johnson} B.,  {Levesque} E.,
  {Berg} D.,  2018, preprint, \href
  {http://adsabs.harvard.edu/abs/2018arXiv180304425B} {} (\mn@eprint {arXiv}
  {1803.04425})

\bibitem[\protect\citeauthoryear{{Calzetti}, {Armus}, {Bohlin}, {Kinney},
  {Koornneef}  \& {Storchi-Bergmann}}{{Calzetti} et~al.}{2000}]{Calzetti2000}
{Calzetti} D.,  {Armus} L.,  {Bohlin} R.~C.,  {Kinney} A.~L.,  {Koornneef} J.,
   {Storchi-Bergmann} T.,  2000, \mn@doi [\apj] {10.1086/308692}, \href
  {http://adsabs.harvard.edu/abs/2000ApJ...533..682C} {533, 682}

\bibitem[\protect\citeauthoryear{{Cerny} et~al.,}{{Cerny}
  et~al.}{2018}]{Cerny2018}
{Cerny} C.,  et~al., 2018, \mn@doi [\apj] {10.3847/1538-4357/aabe7b}, \href
  {http://adsabs.harvard.edu/abs/2018ApJ...859..159C} {859, 159}

\bibitem[\protect\citeauthoryear{{Chabrier}}{{Chabrier}}{2003}]{Chabrier2003}
{Chabrier} G.,  2003, \mn@doi [\pasp] {10.1086/376392}, \href
  {http://cdsads.u-strasbg.fr/abs/2003PASP..115..763C} {115, 763}

\bibitem[\protect\citeauthoryear{{Chevallard} \& {Charlot}}{{Chevallard} \&
  {Charlot}}{2016}]{Chevallard2016}
{Chevallard} J.,  {Charlot} S.,  2016, \mn@doi [\mnras]
  {10.1093/mnras/stw1756}, \href
  {http://adsabs.harvard.edu/abs/2016MNRAS.462.1415C} {462, 1415}

\bibitem[\protect\citeauthoryear{{Christensen} et~al.,}{{Christensen}
  et~al.}{2012}]{Christensen2012}
{Christensen} L.,  et~al., 2012, \mn@doi [\mnras]
  {10.1111/j.1365-2966.2012.22006.x}, \href
  {http://cdsads.u-strasbg.fr/abs/2012MNRAS.427.1953C} {427, 1953}

\bibitem[\protect\citeauthoryear{{Cibirka} et~al.,}{{Cibirka}
  et~al.}{2018}]{Cibirka2018}
{Cibirka} N.,  et~al., 2018, \mn@doi [\apj] {10.3847/1538-4357/aad2d3}, \href
  {http://adsabs.harvard.edu/abs/2018ApJ...863..145C} {863, 145}

\bibitem[\protect\citeauthoryear{{Coe} et~al.,}{{Coe} et~al.}{2019}]{Coe2019}
{Coe} D.,  et~al., 2019, arXiv e-prints, \href
  {http://adsabs.harvard.edu/abs/2019arXiv190302002C} {}

\bibitem[\protect\citeauthoryear{{Curtis-Lake} et~al.,}{{Curtis-Lake}
  et~al.}{2016}]{Curtislake2016}
{Curtis-Lake} E.,  et~al., 2016, \mn@doi [\mnras] {10.1093/mnras/stv3017},
  \href {http://adsabs.harvard.edu/abs/2016MNRAS.457..440C} {457, 440}

\bibitem[\protect\citeauthoryear{{De Barros}, {Oesch}, {Labb{\'e}}, {Stefanon},
  {Gonz{\'a}lez}, {Smit}, {Bouwens}  \& {Illingworth}}{{De Barros}
  et~al.}{2019}]{deBarros2019}
{De Barros} S.,  {Oesch} P.~A.,  {Labb{\'e}} I.,  {Stefanon} M.,
  {Gonz{\'a}lez} V.,  {Smit} R.,  {Bouwens} R.~J.,   {Illingworth} G.~D.,
  2019, \mn@doi [\mnras] {10.1093/mnras/stz940}, \href
  {https://ui.adsabs.harvard.edu/abs/2019MNRAS.tmp..907D} {p.~907}

\bibitem[\protect\citeauthoryear{{Dressler} et~al.,}{{Dressler}
  et~al.}{2011}]{Dressler2011}
{Dressler} A.,  et~al., 2011, \mn@doi [\pasp] {10.1086/658908}, \href
  {http://adsabs.harvard.edu/abs/2011PASP..123..288D} {123, 288}

\bibitem[\protect\citeauthoryear{{Du}, {Shapley}, {Martin}  \& {Coil}}{{Du}
  et~al.}{2017}]{Du2017}
{Du} X.,  {Shapley} A.~E.,  {Martin} C.~L.,   {Coil} A.~L.,  2017, \mn@doi
  [\apj] {10.3847/1538-4357/aa64cf}, \href
  {http://adsabs.harvard.edu/abs/2017ApJ...838...63D} {838, 63}

\bibitem[\protect\citeauthoryear{{Du} et~al.,}{{Du} et~al.}{2018}]{Du2018}
{Du} X.,  et~al., 2018, preprint, \href
  {http://adsabs.harvard.edu/abs/2018arXiv180305912D} {} (\mn@eprint {arXiv}
  {1803.05912})

\bibitem[\protect\citeauthoryear{{Eldridge} \& {Stanway}}{{Eldridge} \&
  {Stanway}}{2016}]{Eldridge2016}
{Eldridge} J.~J.,  {Stanway} E.~R.,  2016, \mn@doi [\mnras]
  {10.1093/mnras/stw1772}, \href
  {http://adsabs.harvard.edu/abs/2016MNRAS.462.3302E} {462, 3302}

\bibitem[\protect\citeauthoryear{{Erb}, {Pettini}, {Shapley}, {Steidel}, {Law}
  \& {Reddy}}{{Erb} et~al.}{2010}]{Erb2010}
{Erb} D.~K.,  {Pettini} M.,  {Shapley} A.~E.,  {Steidel} C.~C.,  {Law} D.~R.,
  {Reddy} N.~A.,  2010, \mn@doi [\apj] {10.1088/0004-637X/719/2/1168}, \href
  {http://cdsads.u-strasbg.fr/abs/2010ApJ...719.1168E} {719, 1168}

\bibitem[\protect\citeauthoryear{{Esteban}, {Garc{\'\i}a-Rojas}, {Carigi},
  {Peimbert}, {Bresolin}, {L{\'o}pez-S{\'a}nchez}  \& {Mesa-Delgado}}{{Esteban}
  et~al.}{2014}]{Esteban2014}
{Esteban} C.,  {Garc{\'\i}a-Rojas} J.,  {Carigi} L.,  {Peimbert} M.,
  {Bresolin} F.,  {L{\'o}pez-S{\'a}nchez} A.~R.,   {Mesa-Delgado} A.,  2014,
  \mn@doi [\mnras] {10.1093/mnras/stu1177}, \href
  {https://ui.adsabs.harvard.edu/abs/2014MNRAS.443..624E} {443, 624}

\bibitem[\protect\citeauthoryear{{Feltre}, {Charlot}  \& {Gutkin}}{{Feltre}
  et~al.}{2016}]{Feltre2016}
{Feltre} A.,  {Charlot} S.,   {Gutkin} J.,  2016, \mn@doi [\mnras]
  {10.1093/mnras/stv2794}, \href
  {http://adsabs.harvard.edu/abs/2016MNRAS.456.3354F} {456, 3354}

\bibitem[\protect\citeauthoryear{{Ferland} et~al.,}{{Ferland}
  et~al.}{2013}]{Ferland2013}
{Ferland} G.~J.,  et~al., 2013, \rmxaa, \href
  {http://cdsads.u-strasbg.fr/abs/2013RMxAA..49..137F} {49, 137}

\bibitem[\protect\citeauthoryear{{Finkelstein} et~al.,}{{Finkelstein}
  et~al.}{2013}]{Finkelstein2013}
{Finkelstein} S.~L.,  et~al., 2013, \mn@doi [\nat] {10.1038/nature12657}, \href
  {http://cdsads.u-strasbg.fr/abs/2013Natur.502..524F} {502, 524}

\bibitem[\protect\citeauthoryear{{Finkelstein} et~al.,}{{Finkelstein}
  et~al.}{2015}]{Finkelstein2015}
{Finkelstein} S.~L.,  et~al., 2015, \mn@doi [\apj]
  {10.1088/0004-637X/810/1/71}, \href
  {http://adsabs.harvard.edu/abs/2015ApJ...810...71F} {810, 71}

\bibitem[\protect\citeauthoryear{{Finkelstein} et~al.,}{{Finkelstein}
  et~al.}{2019}]{Finkelstein2019}
{Finkelstein} S.~L.,  et~al., 2019, arXiv e-prints, \href
  {https://ui.adsabs.harvard.edu/abs/2019arXiv190202792F} {p. arXiv:1902.02792}

\bibitem[\protect\citeauthoryear{{Fletcher}, {Tang}, {Robertson}, {Nakajima},
  {Ellis}, {Stark}  \& {Inoue}}{{Fletcher} et~al.}{2019}]{Fletcher2019}
{Fletcher} T.~J.,  {Tang} M.,  {Robertson} B.~E.,  {Nakajima} K.,  {Ellis}
  R.~S.,  {Stark} D.~P.,   {Inoue} A.,  2019, \mn@doi [\apj]
  {10.3847/1538-4357/ab2045}, \href
  {https://ui.adsabs.harvard.edu/abs/2019ApJ...878...87F} {878, 87}

\bibitem[\protect\citeauthoryear{{Garnett}, {Skillman}, {Dufour}, {Peimbert},
  {Torres-Peimbert}, {Terlevich}, {Terlevich}  \& {Shields}}{{Garnett}
  et~al.}{1995}]{Garnett1995}
{Garnett} D.~R.,  {Skillman} E.~D.,  {Dufour} R.~J.,  {Peimbert} M.,
  {Torres-Peimbert} S.,  {Terlevich} R.,  {Terlevich} E.,   {Shields} G.~A.,
  1995, \mn@doi [\apj] {10.1086/175503}, \href
  {http://adsabs.harvard.edu/abs/1995ApJ...443...64G} {443, 64}

\bibitem[\protect\citeauthoryear{{Gonz{\'a}lez}, {Bouwens}, {Illingworth},
  {Labb{\'e}}, {Oesch}, {Franx}  \& {Magee}}{{Gonz{\'a}lez}
  et~al.}{2014}]{Gonzalez2014}
{Gonz{\'a}lez} V.,  {Bouwens} R.,  {Illingworth} G.,  {Labb{\'e}} I.,  {Oesch}
  P.,  {Franx} M.,   {Magee} D.,  2014, \mn@doi [\apj]
  {10.1088/0004-637X/781/1/34}, \href
  {http://adsabs.harvard.edu/abs/2014ApJ...781...34G} {781, 34}

\bibitem[\protect\citeauthoryear{{Groves}, {Heckman}  \& {Kauffmann}}{{Groves}
  et~al.}{2006}]{Groves2006}
{Groves} B.~A.,  {Heckman} T.~M.,   {Kauffmann} G.,  2006, \mn@doi [\mnras]
  {10.1111/j.1365-2966.2006.10812.x}, \href
  {https://ui.adsabs.harvard.edu/abs/2006MNRAS.371.1559G} {371, 1559}

\bibitem[\protect\citeauthoryear{{Gutkin}, {Charlot}  \& {Bruzual}}{{Gutkin}
  et~al.}{2016}]{Gutkin2016}
{Gutkin} J.,  {Charlot} S.,   {Bruzual} G.,  2016, \mn@doi [\mnras]
  {10.1093/mnras/stw1716}, \href
  {http://adsabs.harvard.edu/abs/2016MNRAS.462.1757G} {462, 1757}

\bibitem[\protect\citeauthoryear{{Hainline}, {Shapley}, {Greene}  \&
  {Steidel}}{{Hainline} et~al.}{2011}]{Hainline2011}
{Hainline} K.~N.,  {Shapley} A.~E.,  {Greene} J.~E.,   {Steidel} C.~C.,  2011,
  \mn@doi [\apj] {10.1088/0004-637X/733/1/31}, \href
  {http://adsabs.harvard.edu/abs/2011ApJ...733...31H} {733, 31}

\bibitem[\protect\citeauthoryear{{Hashimoto} et~al.,}{{Hashimoto}
  et~al.}{2018}]{Hashimoto2018}
{Hashimoto} T.,  et~al., 2018, \mn@doi [\nat] {10.1038/s41586-018-0117-z},
  \href {https://ui.adsabs.harvard.edu/abs/2018Natur.557..392H} {557, 392}

\bibitem[\protect\citeauthoryear{{Hu} et~al.,}{{Hu} et~al.}{2017}]{Hu2017}
{Hu} W.,  et~al., 2017, \mn@doi [\apjl] {10.3847/2041-8213/aa8401}, \href
  {http://adsabs.harvard.edu/abs/2017ApJ...845L..16H} {845, L16}

\bibitem[\protect\citeauthoryear{{Hutchison} et~al.,}{{Hutchison}
  et~al.}{2019}]{Hutchison2019}
{Hutchison} T.~A.,  et~al., 2019, arXiv e-prints, \href
  {https://ui.adsabs.harvard.edu/abs/2019arXiv190508812H} {p. arXiv:1905.08812}

\bibitem[\protect\citeauthoryear{{Izotov} \& {Thuan}}{{Izotov} \&
  {Thuan}}{2008}]{Izotov2008}
{Izotov} Y.~I.,  {Thuan} T.~X.,  2008, \mn@doi [\apj] {10.1086/591660}, \href
  {https://ui.adsabs.harvard.edu/abs/2008ApJ...687..133I} {687, 133}

\bibitem[\protect\citeauthoryear{{Izotov}, {Worseck}, {Schaerer}, {Guseva},
  {Thuan}, {Fricke}  \& {Orlitov{\'a}}}{{Izotov} et~al.}{2018}]{Izotov2018}
{Izotov} Y.~I.,  {Worseck} G.,  {Schaerer} D.,  {Guseva} N.~G.,  {Thuan} T.~X.,
   {Fricke} Verhamme A.,   {Orlitov{\'a}} I.,  2018, \mn@doi [\mnras]
  {10.1093/mnras/sty1378}, \href
  {https://ui.adsabs.harvard.edu/abs/2018MNRAS.478.4851I} {478, 4851}

\bibitem[\protect\citeauthoryear{{Jaskot} \& {Ravindranath}}{{Jaskot} \&
  {Ravindranath}}{2016}]{Jaskot2016}
{Jaskot} A.~E.,  {Ravindranath} S.,  2016, \mn@doi [\apj]
  {10.3847/1538-4357/833/2/136}, \href
  {http://adsabs.harvard.edu/abs/2016ApJ...833..136J} {833, 136}

\bibitem[\protect\citeauthoryear{{Jones}, {Martin}  \& {Cooper}}{{Jones}
  et~al.}{2015}]{Jones2015}
{Jones} T.,  {Martin} C.,   {Cooper} M.~C.,  2015, \mn@doi [\apj]
  {10.1088/0004-637X/813/2/126}, \href
  {http://adsabs.harvard.edu/abs/2015ApJ...813..126J} {813, 126}

\bibitem[\protect\citeauthoryear{{Kauffmann} et~al.,}{{Kauffmann}
  et~al.}{2003}]{Kauffmann2003}
{Kauffmann} G.,  et~al., 2003, \mn@doi [\mnras]
  {10.1111/j.1365-2966.2003.07154.x}, \href
  {http://adsabs.harvard.edu/abs/2003MNRAS.346.1055K} {346, 1055}

\bibitem[\protect\citeauthoryear{{Kelson}}{{Kelson}}{2003}]{Kelson2003}
{Kelson} D.~D.,  2003, \mn@doi [\pasp] {10.1086/375502}, \href
  {http://adsabs.harvard.edu/abs/2003PASP..115..688K} {115, 688}

\bibitem[\protect\citeauthoryear{{Kewley}, {Dopita}, {Sutherland}, {Heisler}
  \& {Trevena}}{{Kewley} et~al.}{2001}]{Kewley2001}
{Kewley} L.~J.,  {Dopita} M.~A.,  {Sutherland} R.~S.,  {Heisler} C.~A.,
  {Trevena} J.,  2001, \mn@doi [\apj] {10.1086/321545}, \href
  {http://adsabs.harvard.edu/abs/2001ApJ...556..121K} {556, 121}

\bibitem[\protect\citeauthoryear{{Labb{\'e}} et~al.,}{{Labb{\'e}}
  et~al.}{2013}]{Labbe2013}
{Labb{\'e}} I.,  et~al., 2013, \mn@doi [\apjl] {10.1088/2041-8205/777/2/L19},
  \href {http://adsabs.harvard.edu/abs/2013ApJ...777L..19L} {777, L19}

\bibitem[\protect\citeauthoryear{{Laporte}, {Nakajima}, {Ellis}, {Zitrin},
  {Stark}, {Mainali}  \& {Roberts-Borsani}}{{Laporte}
  et~al.}{2017}]{Laporte2017}
{Laporte} N.,  {Nakajima} K.,  {Ellis} R.~S.,  {Zitrin} A.,  {Stark} D.~P.,
  {Mainali} R.,   {Roberts-Borsani} G.~W.,  2017, \mn@doi [\apj]
  {10.3847/1538-4357/aa96a8}, \href
  {http://adsabs.harvard.edu/abs/2017ApJ...851...40L} {851, 40}

\bibitem[\protect\citeauthoryear{{Le F{\`e}vre} et~al.,}{{Le F{\`e}vre}
  et~al.}{2019}]{LeFevre2019}
{Le F{\`e}vre} O.,  et~al., 2019, \mn@doi [\aap] {10.1051/0004-6361/201732197},
  \href {https://ui.adsabs.harvard.edu/abs/2019A&A...625A..51L} {625, A51}

\bibitem[\protect\citeauthoryear{{Leitherer}, {Tremonti}, {Heckman}  \&
  {Calzetti}}{{Leitherer} et~al.}{2011}]{Leitherer2011}
{Leitherer} C.,  {Tremonti} C.~A.,  {Heckman} T.~M.,   {Calzetti} D.,  2011,
  \mn@doi [\aj] {10.1088/0004-6256/141/2/37}, \href
  {http://adsabs.harvard.edu/abs/2011AJ....141...37L} {141, 37}

\bibitem[\protect\citeauthoryear{{Livermore}, {Finkelstein}  \&
  {Lotz}}{{Livermore} et~al.}{2017}]{Livermore2017}
{Livermore} R.~C.,  {Finkelstein} S.~L.,   {Lotz} J.~M.,  2017, \mn@doi [\apj]
  {10.3847/1538-4357/835/2/113}, \href
  {http://adsabs.harvard.edu/abs/2017ApJ...835..113L} {835, 113}

\bibitem[\protect\citeauthoryear{{Luridiana}, {Morisset}  \&
  {Shaw}}{{Luridiana} et~al.}{2015}]{Luridiana2015}
{Luridiana} V.,  {Morisset} C.,   {Shaw} R.~A.,  2015, \mn@doi [\aap]
  {10.1051/0004-6361/201323152}, \href
  {https://ui.adsabs.harvard.edu/abs/2015A&A...573A..42L} {573, A42}

\bibitem[\protect\citeauthoryear{{Mainali}, {Kollmeier}, {Stark}, {Simcoe},
  {Walth}, {Newman}  \& {Miller}}{{Mainali} et~al.}{2017}]{Mainali2017}
{Mainali} R.,  {Kollmeier} J.~A.,  {Stark} D.~P.,  {Simcoe} R.~A.,  {Walth} G.,
   {Newman} A.~B.,   {Miller} D.~R.,  2017, \mn@doi [\apjl]
  {10.3847/2041-8213/836/1/L14}, \href
  {http://adsabs.harvard.edu/abs/2017ApJ...836L..14M} {836, L14}

\bibitem[\protect\citeauthoryear{{Mainali} et~al.,}{{Mainali}
  et~al.}{2018}]{Mainali2018}
{Mainali} R.,  et~al., 2018, \mn@doi [\mnras] {10.1093/mnras/sty1640}, \href
  {http://adsabs.harvard.edu/abs/2018MNRAS.479.1180M} {479, 1180}

\bibitem[\protect\citeauthoryear{{Maseda} et~al.,}{{Maseda}
  et~al.}{2014}]{Maseda2014}
{Maseda} M.~V.,  et~al., 2014, \mn@doi [\apj] {10.1088/0004-637X/791/1/17},
  \href {http://adsabs.harvard.edu/abs/2014ApJ...791...17M} {791, 17}

\bibitem[\protect\citeauthoryear{{McLure} et~al.,}{{McLure}
  et~al.}{2013}]{McLure2013}
{McLure} R.~J.,  et~al., 2013, \mn@doi [\mnras] {10.1093/mnras/stt627}, \href
  {http://cdsads.u-strasbg.fr/abs/2013MNRAS.432.2696M} {432, 2696}

\bibitem[\protect\citeauthoryear{{Moll{\'a}}, {Garc{\'{\i}}a-Vargas}  \&
  {Bressan}}{{Moll{\'a}} et~al.}{2009}]{Molla2009}
{Moll{\'a}} M.,  {Garc{\'{\i}}a-Vargas} M.~L.,   {Bressan} A.,  2009, \mn@doi
  [\mnras] {10.1111/j.1365-2966.2009.15160.x}, \href
  {http://adsabs.harvard.edu/abs/2009MNRAS.398..451M} {398, 451}

\bibitem[\protect\citeauthoryear{{Nakajima} et~al.,}{{Nakajima}
  et~al.}{2018}]{Nakajima2018}
{Nakajima} K.,  et~al., 2018, \mn@doi [\aap] {10.1051/0004-6361/201731935},
  \href {http://adsabs.harvard.edu/abs/2018A%26A...612A..94N} {612, A94}

\bibitem[\protect\citeauthoryear{{Ono} et~al.,}{{Ono} et~al.}{2012}]{Ono2012}
{Ono} Y.,  et~al., 2012, \mn@doi [\apj] {10.1088/0004-637X/744/2/83}, \href
  {http://cdsads.u-strasbg.fr/abs/2012ApJ...744...83O} {744, 83}

\bibitem[\protect\citeauthoryear{{Ono} et~al.,}{{Ono} et~al.}{2013}]{Ono2013}
{Ono} Y.,  et~al., 2013, \mn@doi [\apj] {10.1088/0004-637X/777/2/155}, \href
  {http://adsabs.harvard.edu/abs/2013ApJ...777..155O} {777, 155}

\bibitem[\protect\citeauthoryear{{Osterbrock} \& {Ferland}}{{Osterbrock} \&
  {Ferland}}{2006}]{Osterbrock2006}
{Osterbrock} D.~E.,  {Ferland} G.~J.,  2006, {Astrophysics of gaseous nebulae
  and active galactic nuclei}

\bibitem[\protect\citeauthoryear{{Papovich} et~al.,}{{Papovich}
  et~al.}{2019}]{Papovich2019}
{Papovich} C.,  et~al., 2019, in \baas. p.~266 (\mn@eprint {arXiv}
  {1903.04524})

\bibitem[\protect\citeauthoryear{{Paterno-Mahler} et~al.,}{{Paterno-Mahler}
  et~al.}{2018}]{PaternoMahler2018}
{Paterno-Mahler} R.,  et~al., 2018, \mn@doi [\apj] {10.3847/1538-4357/aad239},
  \href {http://adsabs.harvard.edu/abs/2018ApJ...863..154P} {863, 154}

\bibitem[\protect\citeauthoryear{{Peimbert} \& {Peimbert}}{{Peimbert} \&
  {Peimbert}}{2002}]{Peimbert2002}
{Peimbert} M.,  {Peimbert} A.,  2002, in {Claria} J.~J.,  {Garcia Lambas} D.,
  {Levato} H.,  eds,  Revista Mexicana de Astronomia y Astrofisica Conference
  Series Vol. 14, Revista Mexicana de Astronomia y Astrofisica Conference
  Series. pp 47--52 (\mn@eprint {arXiv} {astro-ph/0204087})

\bibitem[\protect\citeauthoryear{{P{\'e}rez-Montero}}{{P{\'e}rez-Montero}}{2017}]{Perez-Montero2017a}
{P{\'e}rez-Montero} E.,  2017, \mn@doi [\pasp] {10.1088/1538-3873/aa5abb},
  \href {https://ui.adsabs.harvard.edu/abs/2017PASP..129d3001P} {129, 043001}

\bibitem[\protect\citeauthoryear{{P{\'e}rez-Montero} \&
  {Amor{\'\i}n}}{{P{\'e}rez-Montero} \&
  {Amor{\'\i}n}}{2017}]{Perez-Montero2017}
{P{\'e}rez-Montero} E.,  {Amor{\'\i}n} R.,  2017, \mn@doi [\mnras]
  {10.1093/mnras/stx186}, \href
  {https://ui.adsabs.harvard.edu/abs/2017MNRAS.467.1287P} {467, 1287}

\bibitem[\protect\citeauthoryear{{Reines}, {Greene}  \& {Geha}}{{Reines}
  et~al.}{2013}]{Reines2013}
{Reines} A.~E.,  {Greene} J.~E.,   {Geha} M.,  2013, \mn@doi [\apj]
  {10.1088/0004-637X/775/2/116}, \href
  {https://ui.adsabs.harvard.edu/abs/2013ApJ...775..116R} {775, 116}

\bibitem[\protect\citeauthoryear{{Rigby}, {Bayliss}, {Gladders}, {Sharon},
  {Wuyts}, {Dahle}, {Johnson}  \& {Pe{\~n}a-Guerrero}}{{Rigby}
  et~al.}{2015}]{Rigby2015}
{Rigby} J.~R.,  {Bayliss} M.~B.,  {Gladders} M.~D.,  {Sharon} K.,  {Wuyts} E.,
  {Dahle} H.,  {Johnson} T.,   {Pe{\~n}a-Guerrero} M.,  2015, \mn@doi [\apjl]
  {10.1088/2041-8205/814/1/L6}, \href
  {http://adsabs.harvard.edu/abs/2015ApJ...814L...6R} {814, L6}

\bibitem[\protect\citeauthoryear{{Robertson} et~al.,}{{Robertson}
  et~al.}{2013}]{Robertson2013}
{Robertson} B.~E.,  et~al., 2013, \mn@doi [\apj] {10.1088/0004-637X/768/1/71},
  \href {http://adsabs.harvard.edu/abs/2013ApJ...768...71R} {768, 71}

\bibitem[\protect\citeauthoryear{{Salmon} et~al.,}{{Salmon}
  et~al.}{2015}]{Salmon2015}
{Salmon} B.,  et~al., 2015, \mn@doi [\apj] {10.1088/0004-637X/799/2/183}, \href
  {http://adsabs.harvard.edu/abs/2015ApJ...799..183S} {799, 183}

\bibitem[\protect\citeauthoryear{{Sanders} et~al.,}{{Sanders}
  et~al.}{2016}]{Sanders2016}
{Sanders} R.~L.,  et~al., 2016, \mn@doi [\apj] {10.3847/0004-637X/816/1/23},
  \href {http://adsabs.harvard.edu/abs/2016ApJ...816...23S} {816, 23}

\bibitem[\protect\citeauthoryear{{Sanders} et~al.,}{{Sanders}
  et~al.}{2019}]{Sanders2019}
{Sanders} R.~L.,  et~al., 2019, arXiv e-prints, \href
  {https://ui.adsabs.harvard.edu/abs/2019arXiv190700013S} {p. arXiv:1907.00013}

\bibitem[\protect\citeauthoryear{{Schmidt} et~al.,}{{Schmidt}
  et~al.}{2017}]{Schmidt2017}
{Schmidt} K.~B.,  et~al., 2017, \mn@doi [\apj] {10.3847/1538-4357/aa68a3},
  \href {http://adsabs.harvard.edu/abs/2017ApJ...839...17S} {839, 17}

\bibitem[\protect\citeauthoryear{{Senchyna} et~al.,}{{Senchyna}
  et~al.}{2017}]{Senchyna2017}
{Senchyna} P.,  et~al., 2017, \mn@doi [\mnras] {10.1093/mnras/stx2059}, \href
  {http://adsabs.harvard.edu/abs/2017MNRAS.472.2608S} {472, 2608}

\bibitem[\protect\citeauthoryear{{Senchyna}, {Stark}, {Chevallard}, {Charlot},
  {Jones}  \& {Vidal Garc{\'{\i}}a}}{{Senchyna} et~al.}{2019}]{Senchyna2019}
{Senchyna} P.,  {Stark} D.~P.,  {Chevallard} J.,  {Charlot} S.,  {Jones} T.,
  {Vidal Garc{\'{\i}}a} A.,  2019, arXiv e-prints, \href
  {http://adsabs.harvard.edu/abs/2019arXiv190401615S} {}

\bibitem[\protect\citeauthoryear{{Shapley}, {Steidel}, {Pettini}  \&
  {Adelberger}}{{Shapley} et~al.}{2003}]{Shapley2003}
{Shapley} A.~E.,  {Steidel} C.~C.,  {Pettini} M.,   {Adelberger} K.~L.,  2003,
  \mn@doi [\apj] {10.1086/373922}, \href
  {http://cdsads.u-strasbg.fr/abs/2003ApJ...588...65S} {588, 65}

\bibitem[\protect\citeauthoryear{{Shapley} et~al.,}{{Shapley}
  et~al.}{2015}]{Shapley2015}
{Shapley} A.~E.,  et~al., 2015, \mn@doi [\apj] {10.1088/0004-637X/801/2/88},
  \href {http://adsabs.harvard.edu/abs/2015ApJ...801...88S} {801, 88}

\bibitem[\protect\citeauthoryear{{Simcoe} et~al.,}{{Simcoe}
  et~al.}{2013}]{Simcoe2013}
{Simcoe} R.~A.,  et~al., 2013, \mn@doi [\pasp] {10.1086/670241}, \href
  {http://adsabs.harvard.edu/abs/2013PASP..125..270S} {125, 270}

\bibitem[\protect\citeauthoryear{{Smit} et~al.,}{{Smit}
  et~al.}{2014}]{Smit2014a}
{Smit} R.,  et~al., 2014, \mn@doi [\apj] {10.1088/0004-637X/784/1/58}, \href
  {http://adsabs.harvard.edu/abs/2014ApJ...784...58S} {784, 58}

\bibitem[\protect\citeauthoryear{{Smit} et~al.,}{{Smit}
  et~al.}{2015}]{Smit2015}
{Smit} R.,  et~al., 2015, \mn@doi [\apj] {10.1088/0004-637X/801/2/122}, \href
  {http://adsabs.harvard.edu/abs/2015ApJ...801..122S} {801, 122}

\bibitem[\protect\citeauthoryear{{Stanway}, {Eldridge}  \& {Becker}}{{Stanway}
  et~al.}{2016}]{Stanway2016}
{Stanway} E.~R.,  {Eldridge} J.~J.,   {Becker} G.~D.,  2016, \mn@doi [\mnras]
  {10.1093/mnras/stv2661}, \href
  {http://adsabs.harvard.edu/abs/2016MNRAS.456..485S} {456, 485}

\bibitem[\protect\citeauthoryear{{Stark}}{{Stark}}{2016}]{Stark2016}
{Stark} D.~P.,  2016, \mn@doi [\araa] {10.1146/annurev-astro-081915-023417},
  \href {http://adsabs.harvard.edu/abs/2016ARA%26A..54..761S} {54, 761}

\bibitem[\protect\citeauthoryear{{Stark}, {Schenker}, {Ellis}, {Robertson},
  {McLure}  \& {Dunlop}}{{Stark} et~al.}{2013}]{Stark2013}
{Stark} D.~P.,  {Schenker} M.~A.,  {Ellis} R.,  {Robertson} B.,  {McLure} R.,
  {Dunlop} J.,  2013, \mn@doi [\apj] {10.1088/0004-637X/763/2/129}, \href
  {http://adsabs.harvard.edu/abs/2013ApJ...763..129S} {763, 129}

\bibitem[\protect\citeauthoryear{{Stark} et~al.,}{{Stark}
  et~al.}{2014}]{Stark2014}
{Stark} D.~P.,  et~al., 2014, \mn@doi [\mnras] {10.1093/mnras/stu1618}, \href
  {http://adsabs.harvard.edu/abs/2014MNRAS.445.3200S} {445, 3200}

\bibitem[\protect\citeauthoryear{{Stark} et~al.,}{{Stark}
  et~al.}{2015a}]{Stark2015a}
{Stark} D.~P.,  et~al., 2015a, \mn@doi [\mnras] {10.1093/mnras/stv688}, \href
  {http://adsabs.harvard.edu/abs/2015MNRAS.450.1846S} {450, 1846}

\bibitem[\protect\citeauthoryear{{Stark} et~al.,}{{Stark}
  et~al.}{2015b}]{Stark2015b}
{Stark} D.~P.,  et~al., 2015b, \mn@doi [\mnras] {10.1093/mnras/stv1907}, \href
  {http://adsabs.harvard.edu/abs/2015MNRAS.454.1393S} {454, 1393}

\bibitem[\protect\citeauthoryear{{Stark} et~al.,}{{Stark}
  et~al.}{2017}]{Stark2017}
{Stark} D.~P.,  et~al., 2017, \mn@doi [\mnras] {10.1093/mnras/stw2233}, \href
  {http://adsabs.harvard.edu/abs/2017MNRAS.464..469S} {464, 469}

\bibitem[\protect\citeauthoryear{{Stefanon} et~al.,}{{Stefanon}
  et~al.}{2019}]{Stefanon2019}
{Stefanon} M.,  et~al., 2019, arXiv e-prints, \href
  {https://ui.adsabs.harvard.edu/abs/2019arXiv190210713S} {p. arXiv:1902.10713}

\bibitem[\protect\citeauthoryear{{Steidel}, {Strom}, {Pettini}, {Rudie},
  {Reddy}  \& {Trainor}}{{Steidel} et~al.}{2016}]{Steidel2016}
{Steidel} C.~C.,  {Strom} A.~L.,  {Pettini} M.,  {Rudie} G.~C.,  {Reddy} N.~A.,
    {Trainor} R.~F.,  2016, \mn@doi [\apj] {10.3847/0004-637X/826/2/159}, \href
  {http://adsabs.harvard.edu/abs/2016ApJ...826..159S} {826, 159}

\bibitem[\protect\citeauthoryear{{Strait} et~al.,}{{Strait}
  et~al.}{2019}]{Strait2019}
{Strait} V.,  et~al., 2019, arXiv e-prints, \href
  {https://ui.adsabs.harvard.edu/abs/2019arXiv190509295S} {p. arXiv:1905.09295}

\bibitem[\protect\citeauthoryear{{Strom}, {Steidel}, {Rudie}, {Trainor},
  {Pettini}  \& {Reddy}}{{Strom} et~al.}{2017}]{Strom2017}
{Strom} A.~L.,  {Steidel} C.~C.,  {Rudie} G.~C.,  {Trainor} R.~F.,  {Pettini}
  M.,   {Reddy} N.~A.,  2017, \mn@doi [\apj] {10.3847/1538-4357/836/2/164},
  \href {https://ui.adsabs.harvard.edu/abs/2017ApJ...836..164S} {836, 164}

\bibitem[\protect\citeauthoryear{{Strom}, {Steidel}, {Rudie}, {Trainor}  \&
  {Pettini}}{{Strom} et~al.}{2018}]{Strom2018}
{Strom} A.~L.,  {Steidel} C.~C.,  {Rudie} G.~C.,  {Trainor} R.~F.,   {Pettini}
  M.,  2018, \mn@doi [\apj] {10.3847/1538-4357/aae1a5}, \href
  {https://ui.adsabs.harvard.edu/abs/2018ApJ...868..117S} {868, 117}

\bibitem[\protect\citeauthoryear{{Tang}, {Stark}, {Chevallard}  \&
  {Charlot}}{{Tang} et~al.}{2019}]{Tang2019}
{Tang} M.,  {Stark} D.~P.,  {Chevallard} J.,   {Charlot} S.,  2019, \mn@doi
  [\mnras] {10.1093/mnras/stz2236}, \href
  {https://ui.adsabs.harvard.edu/abs/2019MNRAS.489.2572T} {489, 2572}

\bibitem[\protect\citeauthoryear{{Tilvi} et~al.,}{{Tilvi}
  et~al.}{2016}]{Tilvi2016}
{Tilvi} V.,  et~al., 2016, \mn@doi [\apjl] {10.3847/2041-8205/827/1/L14}, \href
  {http://adsabs.harvard.edu/abs/2016ApJ...827L..14T} {827, L14}

\bibitem[\protect\citeauthoryear{{Trainor}, {Strom}, {Steidel}  \&
  {Rudie}}{{Trainor} et~al.}{2016}]{Trainor2016}
{Trainor} R.~F.,  {Strom} A.~L.,  {Steidel} C.~C.,   {Rudie} G.~C.,  2016,
  \mn@doi [\apj] {10.3847/0004-637X/832/2/171}, \href
  {https://ui.adsabs.harvard.edu/abs/2016ApJ...832..171T} {832, 171}

\bibitem[\protect\citeauthoryear{{Vanzella} et~al.,}{{Vanzella}
  et~al.}{2016a}]{Vanzella2016}
{Vanzella} E.,  et~al., 2016a, \mn@doi [\apjl] {10.3847/2041-8205/821/2/L27},
  \href {http://adsabs.harvard.edu/abs/2016ApJ...821L..27V} {821, L27}

\bibitem[\protect\citeauthoryear{{Vanzella} et~al.,}{{Vanzella}
  et~al.}{2016b}]{Vanzella2016b}
{Vanzella} E.,  et~al., 2016b, \mn@doi [\apj] {10.3847/0004-637X/825/1/41},
  \href {https://ui.adsabs.harvard.edu/abs/2016ApJ...825...41V} {825, 41}

\bibitem[\protect\citeauthoryear{{Vanzella} et~al.,}{{Vanzella}
  et~al.}{2017}]{Vanzella2017}
{Vanzella} E.,  et~al., 2017, \mn@doi [\apj] {10.3847/1538-4357/aa74ae}, \href
  {https://ui.adsabs.harvard.edu/abs/2017ApJ...842...47V} {842, 47}

\bibitem[\protect\citeauthoryear{{Vanzella} et~al.,}{{Vanzella}
  et~al.}{2020}]{Vanzella2020}
{Vanzella} E.,  et~al., 2020, \mn@doi [\mnras] {10.1093/mnras/stz2286}, \href
  {https://ui.adsabs.harvard.edu/abs/2020MNRAS.491.1093V} {491, 1093}

\bibitem[\protect\citeauthoryear{{Volonteri}, {Reines}, {Atek}, {Stark}  \&
  {Trebitsch}}{{Volonteri} et~al.}{2017}]{Volonteri2017}
{Volonteri} M.,  {Reines} A.~E.,  {Atek} H.,  {Stark} D.~P.,   {Trebitsch} M.,
  2017, \mn@doi [\apj] {10.3847/1538-4357/aa93f1}, \href
  {http://adsabs.harvard.edu/abs/2017ApJ...849..155V} {849, 155}

\bibitem[\protect\citeauthoryear{{Zitrin} et~al.,}{{Zitrin}
  et~al.}{2015}]{Zitrin2015}
{Zitrin} A.,  et~al., 2015, \mn@doi [\apjl] {10.1088/2041-8205/810/1/L12},
  \href {http://adsabs.harvard.edu/abs/2015ApJ...810L..12Z} {810, L12}

\bibitem[\protect\citeauthoryear{{Zitrin} et~al.,}{{Zitrin}
  et~al.}{2017}]{Zitrin2017}
{Zitrin} A.,  et~al., 2017, \mn@doi [\apjl] {10.3847/2041-8213/aa69be}, \href
  {http://adsabs.harvard.edu/abs/2017ApJ...839L..11Z} {839, L11}

\bibitem[\protect\citeauthoryear{{de Barros} et~al.,}{{de Barros}
  et~al.}{2016}]{deBarros2016}
{de Barros} S.,  et~al., 2016, \mn@doi [\aap] {10.1051/0004-6361/201527046},
  \href {http://adsabs.harvard.edu/abs/2016A%26A...585A..51D} {585, A51}

\bibitem[\protect\citeauthoryear{{van der Wel} et~al.,}{{van der Wel}
  et~al.}{2011}]{vanderwel2011}
{van der Wel} A.,  et~al., 2011, \mn@doi [\apj] {10.1088/0004-637X/742/2/111},
  \href {http://adsabs.harvard.edu/abs/2011ApJ...742..111V} {742, 111}

\makeatother
\end{thebibliography}

\label{lastpage}
\end{document}